\documentclass[12pt,preprint]{aastex}
\begin{document}
\shorttitle{$T_{90}$ Distribution of GBM GRBs} \shortauthors{Qin et al. }
\title{A Comprehensive Analysis of {\em Fermi} Gamma-ray Burst Data:  III. Energy-Dependent $T_{90}$ Distributions of GBM GRBs and Instrumental Selection Effect on Duration Classification}
\author{Ying Qin\altaffilmark{1}, En-Wei Liang\altaffilmark{1,2,3}, Yun-Feng Liang\altaffilmark{1}, Shuang-Xi Yi\altaffilmark{4}, Lin Lin\altaffilmark{5}, Bin-Bin Zhang\altaffilmark{6}, Jin Zhang\altaffilmark{2,7}, Hou-Jun L\"{u}\altaffilmark{3}, Rui-Jing Lu\altaffilmark{1}, Lian-Zhong L\"{u}\altaffilmark{1}, and Bing Zhang\altaffilmark{3}}
\altaffiltext{1}{Department of Physics and GXU-NAOC Center for Astrophysics and
Space Sciences, Guangxi University, Nanning 530004, China; lew@gxu.edu.cn}
\altaffiltext{2}{The National Astronomical Observatories, Chinese Academy of
Sciences, Beijing 100012, China}
\altaffiltext{3}{Department of Physics and Astronomy, University of Nevada, Las Vegas, NV 89154; zhang@physics.unlv.edu}
\altaffiltext{4}{College of Astronomy and Space Sciences, Nanjing University, Nanjing, 210093, China}
\altaffiltext{5}{Sabanc\i~University, Faculty of Engineering and Natural Sciences, Orhanl\i$-$ Tuzla, \.{I}stanbul 34956, Turkey}
\altaffiltext{6}{Department of Astronomy and Astrophysics, Pennsylvania State University, University Park, PA 16802, USA}
\altaffiltext{7}{College of Physics and Electronic Engineering, Guangxi Teachers Education
University, Nanning, 530001, China}

\begin{abstract}
The durations ($T_{90}$) of 315 GRBs detected with {\em Fermi}/GBM (8-1000 keV) by 2011 September are calculated using the Bayesian Block method. We compare the $T_{90}$ distributions between this sample and those derived from previous/current GRB missions. We show that the $T_{90}$ distribution of this GRB sample is bimodal, with a statistical significance level being comparable to those derived from the {\em BeppoSAX}/GRBM sample and the {\em Swift}/BAT sample, but lower than that derived from the {\em CGRO}/BATSE sample. The short-to-long GRB number ratio is also much lower than that derived from the BATSE sample, i.e., 1:6.5 vs 1:3. We measure $T_{90}$ in several bands, i.e., 8-15, 15-25, 25-50, 50-100, 100-350, and 350-1000 keV, to investigate the energy-dependence effect of the bimodal $T_{90}$ distribution. It is found that the bimodal feature is well observed in the 50-100 and 100-350 keV bands, but is only marginally acceptable in the 25-50 keV and 350-1000 keV bands. The hypothesis of the bimodality is confidently rejected in the 8-15 and 15-25 keV bands. The $T_{90}$ distributions in these bands are roughly consistent with those observed by missions with similar energy bands. The parameter $T_{90}$ as a function of energy follows ${\bar T}_{90}\propto E^{-0.20\pm 0.02}$ for long GRBs. Considering the erratic X-ray and optical flares, the duration of a burst would be even much longer for most GRBs. Our results, together with the observed extended emission of some short GRBs, indicate that the central engine activity time scale would be much longer than $T_{90}$ for both long and short GRBs and the observed bimodal $T_{90}$ distribution may be due to an instrumental selection effect.
\end{abstract}
\section{Introduction}
The gamma-ray burst (GRB) survey with Burst And Transient Source Experiment (BATSE) on board Compton Gamma-Ray Observatory (CGRO) revealed a clear bimodal distribution of the burst duration parameter $T_{90}$, which is measured with the time interval from $5\%$ to $95\%$ accumulative photon counts from the source, and two groups of GRBs, i.e., long vs. short GRBs with a division line at $T_{90}=2$ seconds, was identified (Kouveliotou et al. 1993). It has been long theoretically speculated that long GRBs (LGRBs) are related to the deaths of massive stars (Colgate 1974; Woosley 1993). This is observationally confirmed with the discoveries of the supernovae (SNe) association with some nearby LGRBs (for reviews, see Zhang\ \& M\'{e}szaros 2004; Piran 2005; M\'{e}szaros 2006; Woosley \& Bloom 2006). The rapid localization capacity of the {\em Swift} GRB mission (Gehrels et al. 2004) led to redshift measurements and host galaxy detections of short GRBs (SGRBs). Some nearby SGRBs are found in elliptical/early-type galaxies with very low star formation rates (Gehrels et al. 2005; Bloom et al. 2006; Barthelmy et al. 2005a; Berger et al. 2005) or in the regions with a low star formation rate in star-forming galaxies (Covino et al. 2006; Fox et al. 2005). These are consistent with a non-massive star origin of SGRBs, probably related to the mergers of two compact objects (e.g., Paczy\'nski 1986, 1991; Eichler et al. 1989; Narayan et al. 1992; Bloom et al. 1999; see Nakar 2007 for a review). The SGRB central engine in this model is a hot and dense torus of $0.01\sim 0.3$ M$_\odot$ that is accreted onto a stellar mass black hole. The life time of a SGRB is expected to be typically shorter than 2 seconds (Popham,Woosley \& Fryer, 1999; Narayan, Piran \& Kumar, 2001; Di Matteo, Perna \& Narayan, 2002). This is roughly consistent with the typical $T_{90}$ s of SGRBs observed with BATSE.

It was surprising that two nearby LGRBs, i.e. GRB 060614 ($T_{90}=103$ seconds at $z=0.125$) and 060605 ($T_{90}=4$ seconds at $z=0.089$), had no detection of an accompanied SN, different from other known nearby LGRBs, such as GRB 980425/SN 1998bw (Galama et al. 1998; Kulkarni et al. 1998), GRB 030329/SN 2003dh (Stanek et al. 2003; Hjorth et al. 2003), GRB 031203/SN 2003lw (Malesani et al. 2004), GRB 060218/SN 2006aj (Modjaz et al. 2006; Pian et al. 2006; Sollerman et al. 2006; Mirabal et al. 2006; Cobb et al. 2006), and 100316D/SN 2010bh (Starling et al. 2011; Fan et al. 2011). The prompt emission and afterglow properties of GRB 060614 are similar to those of some nearby ``short'' GRBs that have a non-massive star origin (e.g. Gehrels et al. 2006; Zhang et al. 2007; Gal-Yam et al. 2006). This led to confusion of the long-short classification scheme and a new classification scheme, i.e. Type II (massive star origin) vs. Type I (compact star origin) was proposed (Zhang 2006; Zhang et al. 2007). Note that some well-known Type I GRBs, such as GRB 050724 (Barthelmy et al. 2005; Tanvir et al. 2005; Berger et al. 2005) and 050709 (Hjorth et al. 2005; Villasenor et al. 2005; Fox et al. 2005) also have a long-lasting extended emission component, making $T_{90}$ of these bursts to be $\sim 100$ seconds. In fact, a handful of Type I GRBs show such a component in their lightcurves (Norris et al. 2008; Lin et al. 2008; Zhang et al. 2009). GRB 060614-like nearby LGRBs are therefore likely Type I GRBs with a long, soft extended emission tail, similar to that observed in SGRB 050724 (Zhang et al. 2007). It is also interesting to note that some short-duration GRBs are likely originated from collapse of massive stars (Virgili et al. 2011; Zhang et al. 2009), such as GRB 090426 ($T_{90}=1.28$ in the 15-350 keV band; Levesque et al. 2010a, b; Xin et al. 2010). Some high redshift GRBs, such as GRBs 080913 ($z= 6.7$; Greiner et al. 2009) and 090423 ($z=8.3$; Salvaterra et al. 2009; Tanvir et al. 2009), have a rest-frame short duration ($T_{90}/(1+z) < 2$ seconds), but they share a lot of common properties with LGRBs, and are likely from collapse of massive stars (Zhang et al. 2009; Belczynski et al. 2010; Levesque et al. 2010a; Lin et al. 2010). These observations indicate that the long vs. short GRB classification scheme does not always match the physical Type II vs. Type I classification scheme.  L\"{u} et al. (2010) proposed a new observational parameter defined with the burst isotropic gamma-ray energy and the photon energy of the $\nu f_{\nu}$ spectral peak ($E_{\rm p}$). Similarly, the ratio of the gamma-ray fluence to $E_{\rm p}$ is also suggested as a parameter for GRB classification (Goldstein et al. 2010).

As discussed above, $T_{90}$ is not always a good parameter to conduct GRB classification. It is essential to understand whether the observed bimodal $T_{90}$ distribution is intrinsic or just due to an instrumental selection effect. This is critical for GRB classification and theoretical modeling of GRB progenitors and central engines. Broadband energy band observations with the {\em Fermi} mission not only reveal the spectral components and their temporal evolution (Zhang et al. 2011, Lu et al. 2012; Paper I and II of this series), but also provide an opportunity to study the energy dependence of burst duration and possible instrumental selection effect on the $T_{90}$ distribution. In this paper, we present a detailed analysis on the $T_{90}$ distribution of {\em Fermi}/GBM sample (in Section 2), and compare it with the $T_{90}$ distributions derived from the previous GRB missions (in Section 3). We next explore the instrumental selection effect on $T_{90}$ distribution and energy dependence of $T_{90}$ (in Section 4). We also present discussion on the burst duration by considering late X-ray and optical flares (in Section 5). We show that the bimodal distribution of $T_{90}$ is likely due to an instrumental selection effect and the life time of the central engines of Type I GRBs is essentially longer than 2 seconds for most cases (in Section 6).

\section{Data Reduction and Calculation of $T_{90}$}
We include all 315 GRBs detected by GBM, as reported by the GBM team in GCN circulars up to Sep. 2011. We download the data from {\em Fermi} Archive available at $ftp://legacy.gsfc.nasa.gov/fermi/data/gbm/bursts/$.  The time tagged event (TTE) data have excellent time resolution of 2$\mu$s. The TTE data of the most illuminated NaI detector for each GRB is used for our analysis. The Rmfit(v3.7) package is used for data reduction.

\begin{figure*}
\centering
\includegraphics[angle=0,scale=0.55]{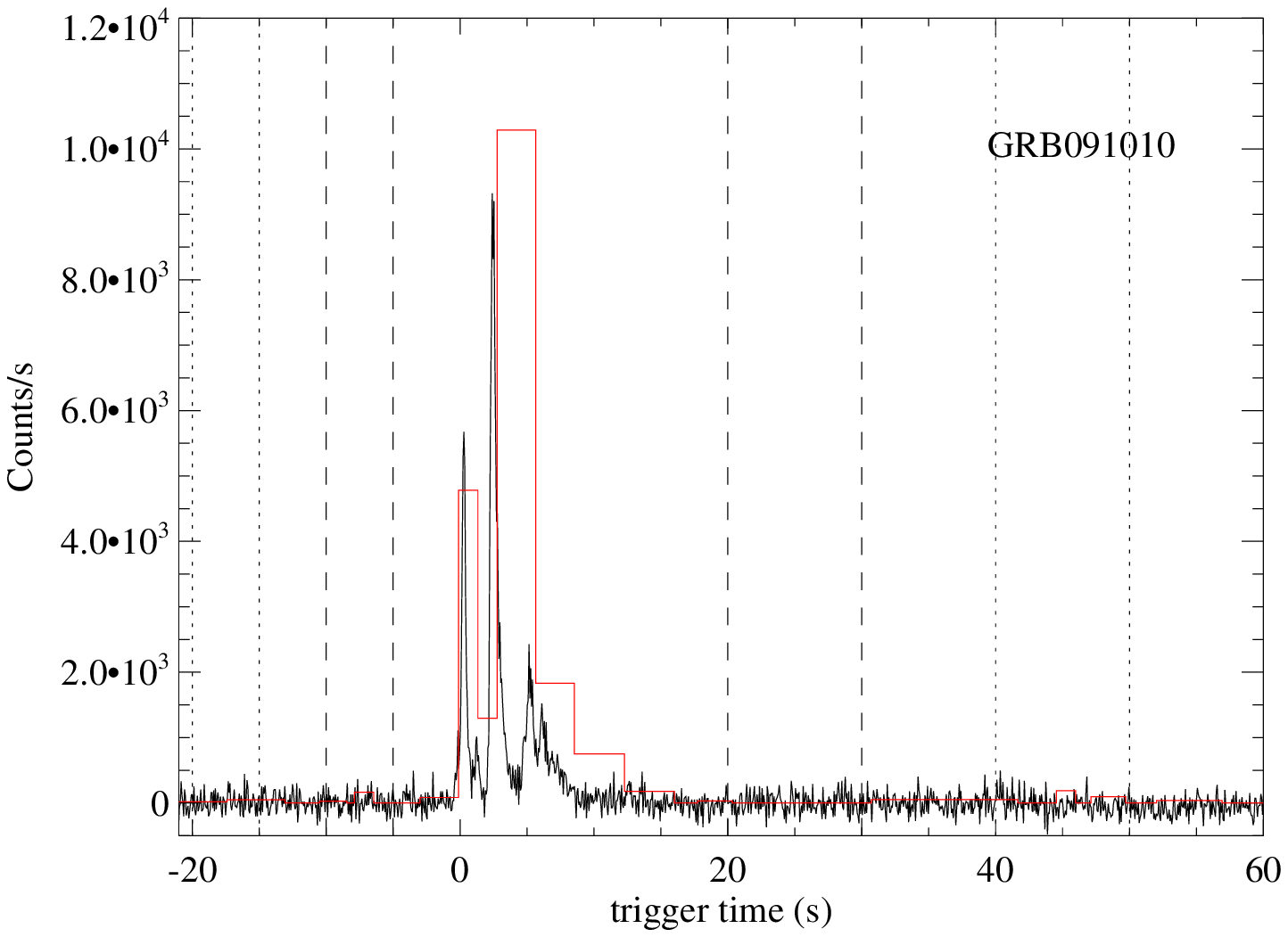}
\includegraphics[angle=0,scale=0.55]{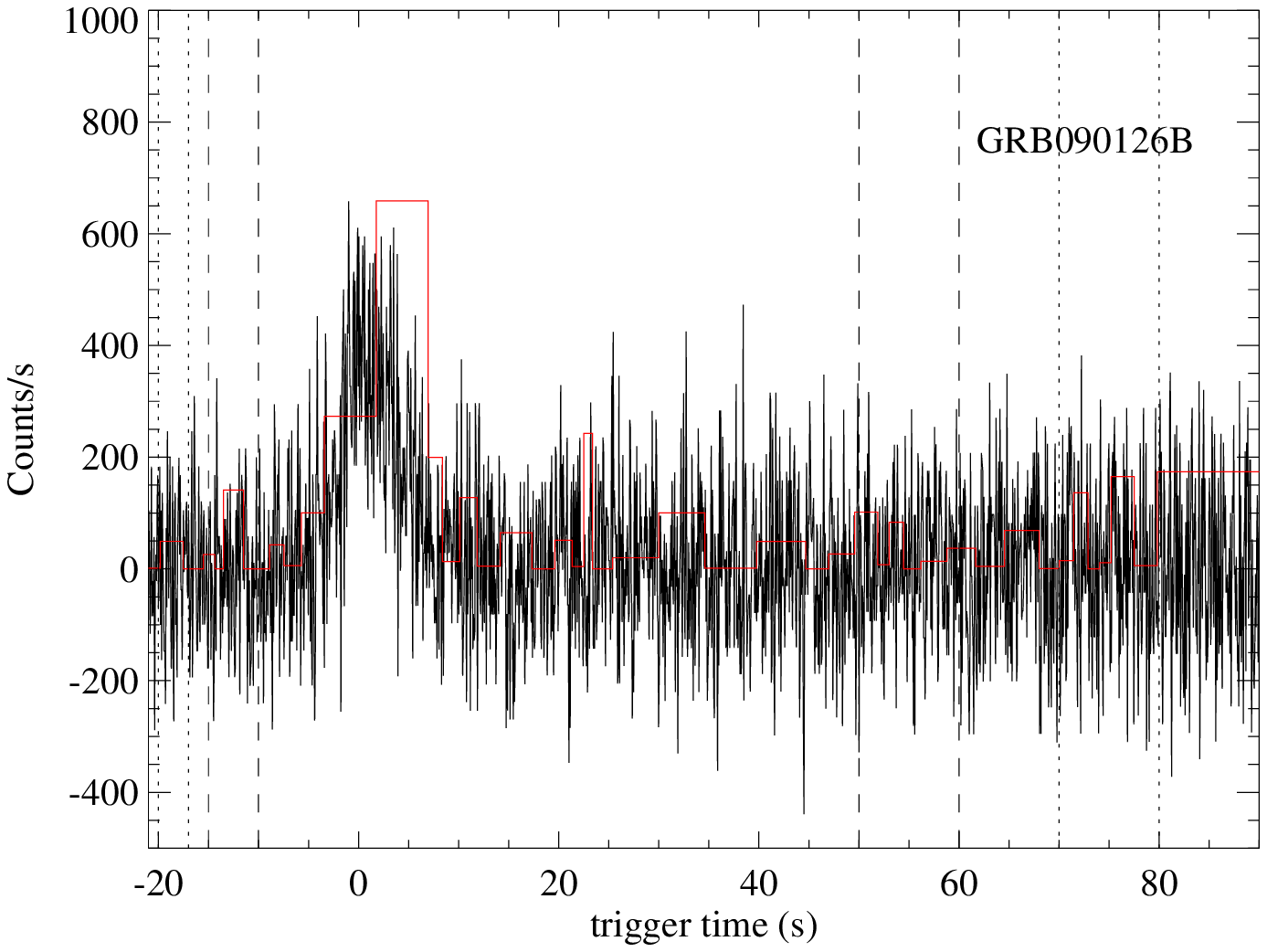}
\includegraphics[angle=0,scale=0.55]{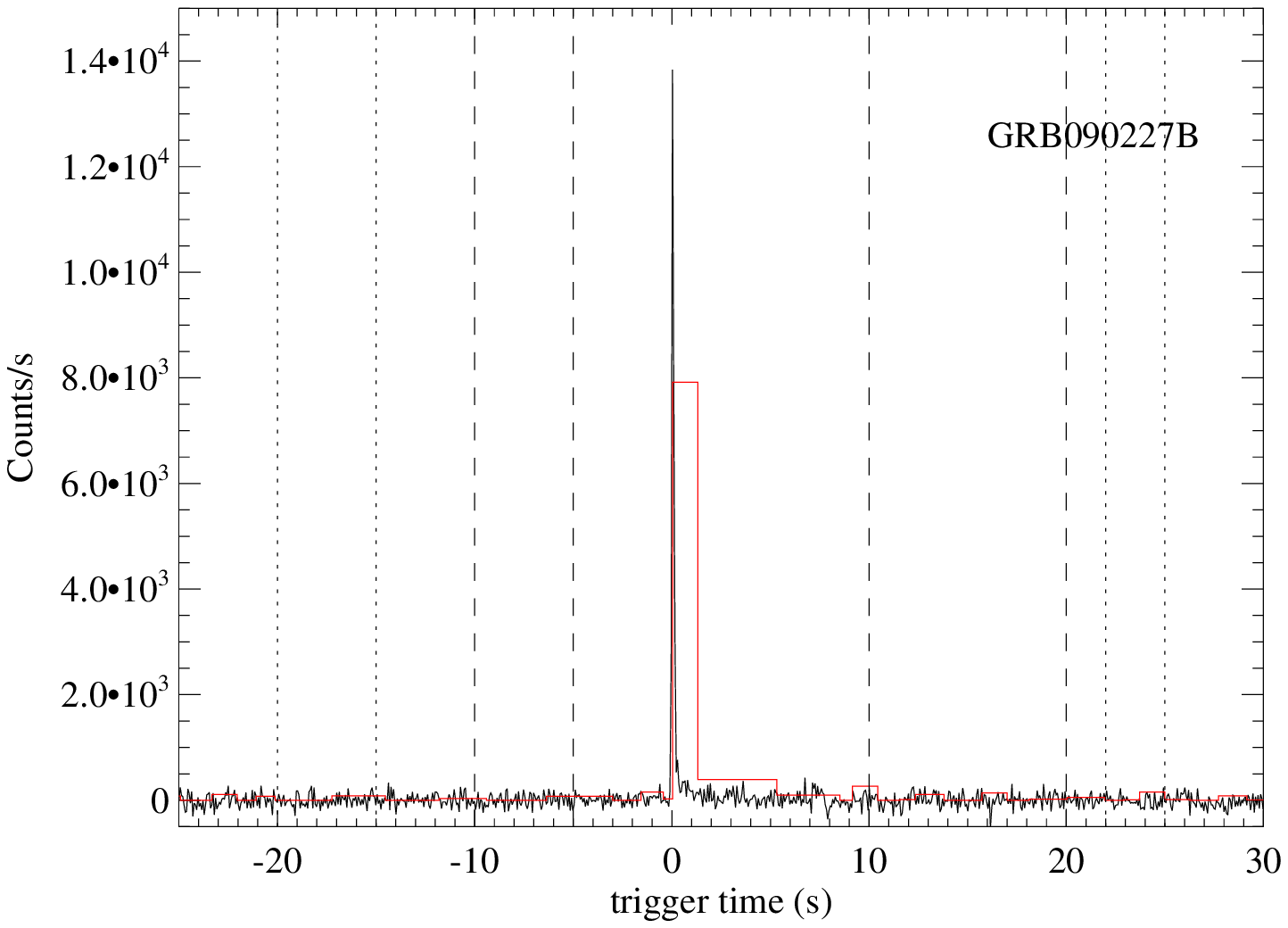}
\caption{Illustrations of background selections (dashed and dotted lines) and the Bayesian blacks (solid lines) for a bright burst(GRB 090910), a  weak burst(GRB 090126B), and a short burst(GRB 090227B). }\label{GRB lightcurves}
\end{figure*}

We select two time intervals that are far before and far after the main burst as background and extract it during the burst phase by a linear fit\footnote{Higher order polynomial fits for background subtraction was also tried, but the derived $T_{90}$ is generally consistent with that derived from the linear fit within error.},  then calculate $T_{90}$ using the Beyasian Block method (Scargle et al. 1998). Note that we do not adopt fixed intervals prior to or post the GRB trigger as the background, since some GRBs have significant emission prior to the trigger, while some others may have a long tail after the main burst. Therefore, the background intervals were visually selected by eye for each GRB. The background subtraction of the lightcurve alters the prior assumption used in the Bayesian Block method and adds additional error into the duration estimation due to the propagation of error from the background. However, it is found that this effect does not significantly affect our $T_{90}$ estimation within error. To clarify the influence of interval selection for background subtraction in our calculation of $T_{90}$, we compare the derived $T_{90}$ by selecting different background intervals for three typical GRBs. The lightcurves and selection of background intervals of these GRBs are displayed in Figure \ref{GRB lightcurves}. For GRB 091010, a bright burst, the derived $T_{90}$ values are $7.616\pm 0.580$ s and $7.552\pm 0.516$ for two different selections of the time intervals for background subtraction, as shown in Figure \ref{GRB lightcurves}. For GRB 090126B, a weak burst, we get $8.032\pm 1.154$ s and $7.968\pm 1.111$s, respectively. For GRB 090227B, a short GRB, we get $1.248\pm 0.601$s and $1.184\pm 0.544$ s, respectively. One can find that the derived $T_{90}$ values are consistent with each other within error for different background selections.

For each GRB in our sample, we extract the 64-ms binned light curves from the TTE data and subtract background for each energy band. We then calculate $T_{90}$ with the Beyasian Block method (Scargle et al. 1998) for each lightcurve. Examples of lightcurve structure obtained from this method are also shown in Figure 1. With this method, we derive the epochs of $t_{5}$ and $t_{95}$, where $t_{5}$ and $t_{95}$ are the times when $5\%$ and $95\%$ of the total count fluence are collected, respectively\footnote{Our calculation is done purely in count space. Since GBM is constantly slewing in orbit, this method could skew the $T_{90}$ estimation of long GRBs.}. In order to reduce fluctuation of $t_{5}$ and $t_{95}$ from a real lightcurve and estimate their error, we generate a sample of $10^3$ mock lightcurve assuming that the error of lightcurve data has a Poisson distribution. The $t_{5}$ and $t_{95}$ values as well as their errors (1$\sigma$) are obtained from a Gaussian fit to their distributions from the mock lightcurve sample. Hence, we get $T_{90}=t_{95}-t_{5}$ and its error $\delta T_{90}=(\delta t_{95}^2+\delta t^2_{5})^{1/2}$. The derived $T_{90}$ are reported in Table 1\footnote{Table 1 is available only in the electronic version}. Note that some lightcurves are too weak to calculate the values of $t_{5}$ and $t_{95}$ with the Bayesian Block method. Therefore, their $T_{90}$ values are not available. 

\begin{figure*}
\centering
\includegraphics[angle=0,scale=0.5]{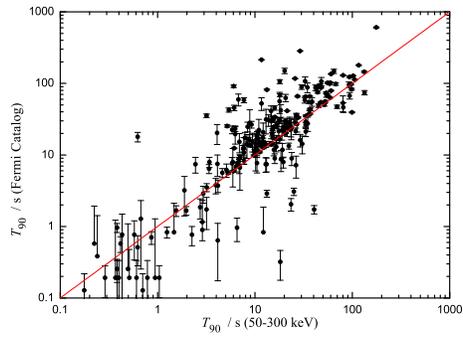}
\caption{Comparison of the derived $T_{90}$ from our method to that reported in the first GBM Catalog (Paciesas et al. 2012) in the 50-300 keV band. The line is the equality line.}\label{T90_comparison}
\end{figure*}

The $T_{90}$ reported in Fermi GBM Catalog by the GBM team is calculated by accumulating the photon fluence through the duration of the burst (Paciesas et al. 2012). In this method, a GRB is split into some time bins and the spectrum of each bin is fit with a model. The photon fluence from the best fit spectral model for each time bin is accumulated to calculate the $T_{90}$. This procedure factors in the detector response and the fact that the angle between the detectors and source are constantly changing. Figure \ref{T90_comparison} compares the derived $T_{90}$ in the 50-300 keV band from the the Bayesian Block method with that derived from the the photon fluence method as reported by Paciesas et al. (2012). It is found that they are generally consistent with each other.

\section{Comparison of the $T_{90}$ Distribution to Other GRB Missions}
\begin{figure*}
\centering
\includegraphics[angle=0,scale=0.6]{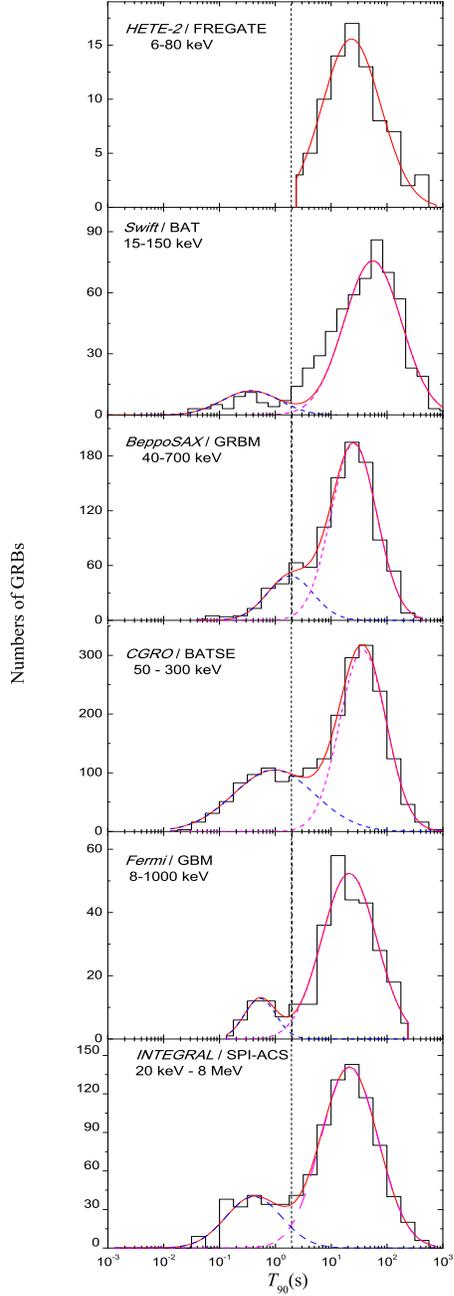}
\caption{Comparison of the $T_{90}$ distributions observed with different instruments. The data of {\em HETE-2/}FREGATE, {\em BeppoSAX}/GRBM, {\rm CGRO}/BATSE, {\rm Swift}/BAT, {\em INTEGRAL/}SPI-ACS, are taken from P\'{e}langeon et al. (2008),  Frontera et al.(2009), Paciesas et al. (1999), Sakamoto et al. (2011), and Savchenko et al. (2012) respectively. The vertical dotted line marks $T_{90}=2$ seconds. The fits to the distributions with two Gaussian functions or one Gaussian function are also shown.}\label{T90_Instruments}
\end{figure*}

Since 1990, GRB surveys in different energy bands have been done with {\em CGRO}/BATSE (50-300 keV), {\em HETE-2}/FREGATE (6-80 keV), {\em BeppoSAX}/GRBM (40-700 keV), {\em Swift}/BAT (15-150 keV), and {\em INTEGRAL}/SPI-ACS (20 keV-8 MeV). The GBM roughly covers the energy bands of these instruments. In Figure \ref{T90_Instruments} and Table 2, we compare the $T_{90}$ distribution of our GBM sample in the 8-1000 keV band with those derived from the data collected by these missions. The data of {\em HETE-2/}FREGATE, {\em BeppoSAX}/GRBM, {\rm CGRO}/BATSE, {\em Swift}/BAT, {\em INTEGRAL/}SPI-ACS, are taken from P\'{e}langeon et al. (2008),  Frontera et al. (2009), Paciesas et al. (1999), Sakamoto et al. (2011), and Savchenko et al. (2012) respectively. Notice that the methods of $T_{90}$ calculations adopted by these instrumental teams may not be exactly the same. In the {\em BeppoSAX}, {\em HETE}-2, {\em CGRO}/BATSE and {\em Integral} samples, the $T_{90}$ values were derived by using the accumulated count rate. The $T_{90}$ values of the {\em Swift} GRB sample were calculated using the Bayesian block method. The $T_{90}$ values of {\em Fermi} GRBs reported by {\em Fermi} GBM team are calculated with the the photon fluence method by considering the instrument response effect, as described in \S2. The derived $T_{90}$ values with different methods for these instruments may have a small systematic bias, which does not greatly affect our comparison of the $T_{90}$ distributions in our analysis.

As shown in Figure \ref{T90_Instruments}, the $T_{90}$ distributions of the LGRB groups observed with different missions are generally consistent with each other, but those of the SGRBs are dramatically different. The {\em HETE}-2 sample even does not have any GRB with $T_{90}<2$ seconds. We fit the $T_{90}$ distributions with a model of two log-normal functions and find that the bimodal distribution feature is revealed only in the BATSE, GBM, {\em BeppoSAX} and {\em INTEGRAL} samples. For the BATSE sample, the $T_{90}$ distribution peaks at $\log T_{90}=-0.02\pm 0.73$ and $\log T_{90}=1.57\pm 0.41$. For the GBM sample, the peaks are at $\log T_{90}=-0.27\pm 0.28$ and $\log T_{90}=1.32\pm 0.49$, respectively. For the {\em BeppoSAX} sample, the peaks are at $\log T_{90}=0.27\pm 0.41$ and $\log T_{90}=1.41\pm 0.40$, respectively. For the {\em INTEGRAL} sample, the peaks are at $\log T_{90}=-0.29\pm 0.06$ and $\log T_{90}=1.36\pm 0.01$, respectively. For the {\em Swift}/BAT sample, two Gaussian components are fit to the data, i.e., $\log T_{90}=-0.45\pm 0.14$ and $\log T_{90}=1.66\pm 0.03$. For the HETE-2 sample, one gets $\log T_{90}=1.36\pm 0.50$. Apparently, the bimodal feature is observed only in the {\em CGRO}/BATSE and {\em Fermi}/GBM-NaI samples, and it is consistent with the result of Zhang et al. (2012), who measure $T_{90}$ with the observed fluence. The short-to-long GRB number ratios are also dramatically different, as reported in Table 2. In the {\em HETE-2}/FREGATE sample, no GRB with $T_{90}<2$ seconds was detected \footnote{Some {\em HETE-2} GRBs are possibly associated with short GRBs in the BATSE band. However, their $T_{90}$ are essentially long in the {\em HETE-2} band. There include GRBs 020531 ($\sim 2$ s in 2-10 keV; Lamb et al. 2003), 040802 ($T_{90}=3.05\pm 0.18$ in 6-80 keV), 050709 (143 s in 15-150 keV; Barthelmy et al. 2005), 051211 ($T_{90}=4.9\pm 0.61$; P\'{e}langeon et al. 2008), and 060121 ($T_{90}=2.6\pm 0.1$; P\'{e}langeon et al. 2008).}. The short-to-long GRB ratios in the Swift/BAT, BeppoSAX/GRBM and {\em INTEGRAL}/SPI-ACS samples are 51:557 (1:11), 111:892 (1:8), and 195:724 (1:3.7), respectively. This ratio becomes 39:253 (1:6.5) in the GBM sample and 500:1541 (1:3) in the {\em CGRO}/BATSE sample.

Since the visible bimodal distribution feature depends on the bin size selection, we further test the bimodality of the $T_{90}$ distribution with the KMM algorithm (Ashman et al. 1994), which is a Gaussian mixture model with parameters estimated by maximum likelihood estimation using the Expectation-Maximization algorithm. This algorithm has been widely used for population and classification studies by astronomers (e.g., Knigge et al. 2011; L\"{u} et al. 2010). The details of the mixture modeling approach and clustering analysis was presented in McLachlan \& Basford (1988) and McLachlan \& Peel (2000). We use the code by Ashman et al. (1994) who applied this technique for detecting and measuring the bimodality of astronomical data sets. The resulting significance of bimodality from the code is a P-value ($P^{\rm KMM}$) and a likelihood ratio of 1 vs 2 Gaussians. The P-value represents the probability of determining the likelihood ratio test statistics from a distribution drawn from a single Gaussian. Notice that the ratio of the likelihood may not be useful without also considering the increased number of parameters in the latter model. Thus, it is most common to use penalized likelihood ratio tests such as the the Bayesian information criterion  (Schwarz et al. 1978) rather than the classical likelihood ratio test. As a result, we only report the $p^{\rm KMM}$ values. A small $P^{\rm KMM}$ rejects the one single Gaussian distribution for the data with higher confidence. Conventionally, $P^{\rm KMM}<0.05$ rejects one Gaussian component in the data. We report the $p^{\rm KMM}$ values for the $T_{90}$ distributions of these instruments in Table 3. The significance level of the bimodal feature in the GBM-NaI sample is comparable to that of the {\em BeppoSAX}/GRBM and {\em Swift}/BAT samples, but is much lower than the {\em CGRO}/BATSE sample. The hypothesis of a bimodal $T_{90}$ distribution for the HETE-2 sample is confidently rejected.

\begin{figure*}
\centering
\includegraphics[angle=0,scale=0.5]{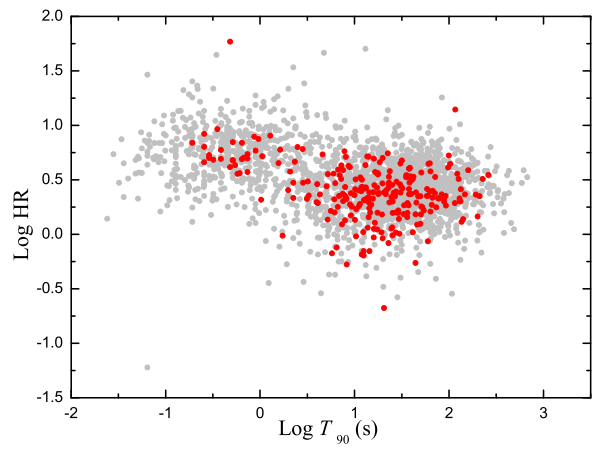}
\caption{The hardness ratio defined with the fluence in the 100-350 keV band to that in the 50-100 keV band as a function of $T_{90}$ for GBM (red) and BATSE (gray) GRBs.}\label{T90_HR}
\end{figure*}

It is also found that the spectra of SGRBs tend to be harder than LGRBs, where the spectral hardness ratio ($HR$) is defined with the fluence ratio between in the 100-350 keV band to that in the 25-50 keV band of BATSE (Kouveliotou et al. 1993). The GBM-NaI energy band covers a similar energy band as that of BATSE, but extends to a lower energy band. Therefore, we also derive the observed fluence in the two energy bands with the spectral parameters reported in GCN circulars for GBM GRBs in our sample, and examine the long-soft vs short-hard classification scheme. Figure \ref{T90_HR} shows the comparison of the GRBs in our sample with the BATSE GRB sample in the $HR-T_{90}$ plane. One can observe that they are consistent with each other.

\section{Energy Dependence of $T_{90}$}
\begin{figure*}
\centering
\includegraphics[angle=0,scale=0.35]{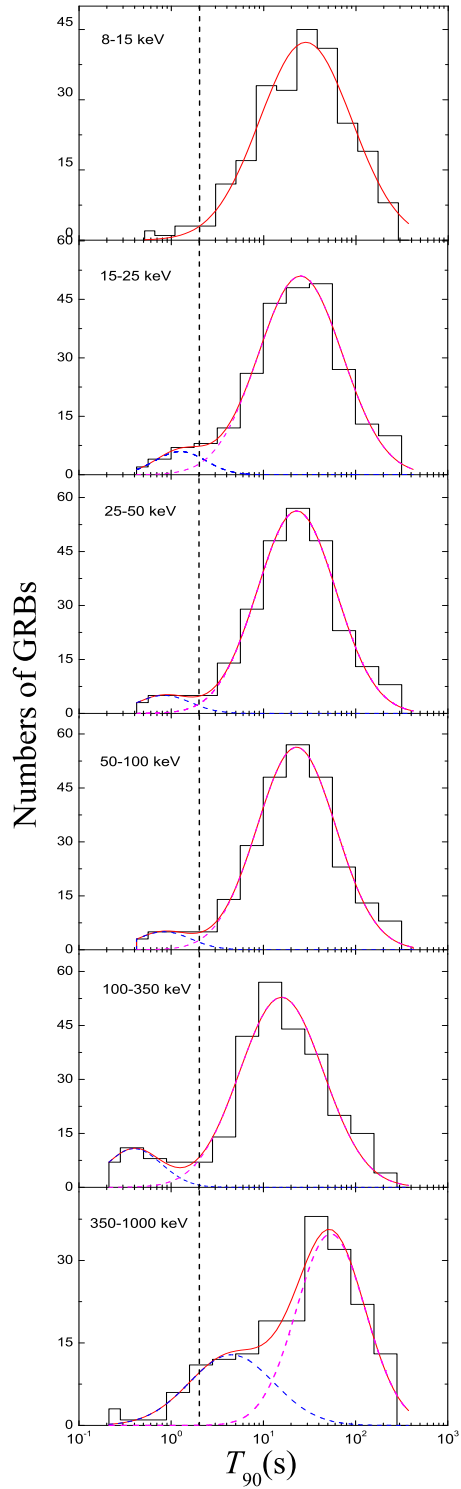}
\caption{ $T_{90}$ distributions in different energy bands. The dotted line corresponds to $T_{90}=2$ seconds. The fits with a model of two log-normal function are also shown (dashed and dotted lines).}\label{T90_bands}
\end{figure*}

As shown above, the $T_{90}$ distribution is instrument-dependent. The deficit of SGRBs in samples observed with instruments in a lower energy band, such as {\em HETE-2}/FREGATE and {\em Swift}/BAT, may be understood as combination of the following two effects. First, since SGRBs are typically harder, one has a lower trigger probability with a soft instrument. Second, many SGRBs have longer softer tails, which are readily detectable in softer detectors. As a result, some GRBs that would be classified as "short" are detected as "long" in soft detectors. As seen in GRB 050724 and 050709, the $T_{90}$ of Type I GRBs is energy-dependent and it could be much longer than 2 seconds. In this section, we investigate the energy dependence of $T_{90}$ with the GBM data. We derive the $T_{90}$ values in the following energy bands: 8-15 keV, 15-25 keV, 25-50 keV, 50-100 keV, 100-350 keV and 350-1000 keV. The short-to-long number ratios in each energy band are also reported in Table 2. The $T_{90}$ distributions are shown in Figure \ref{T90_bands}. Comparing the $T_{90}$ distributions with those observed by other instruments with similar energy bands, it is found that they are roughly consistent with each other, i.e., 8-15 keV band {vs.} {\em HETE-2}/FREGATE (6-80 keV), 15-25 and 25-50 keV bands {vs.} {\em Swift}/BAT (15-150 keV), 50-100 KeV band {vs.} {\em BeppoSAX}/GRBM (40-700 keV), and 100-350 band {vs.} {\em CGRO}/BATSE (25-2000 keV). We also try to fit the $T_{90}$ distributions with the two log-normal component models. Similar results as shown in Figure \ref{T90_Instruments} are obtained. The most significant bimodal $T_{90}$ distribution is seen in the 100-350 keV band. We examine the bimodal feature in the $T_{90}$ distributions of these energy bands with the KMM algorithm and our results are also reported in Table 2. It is found that the bimodal hypothesis is rejected for the $T_{90}$ distributions in the 8-15 and 15-25 keV bands, similar to that observed with {\em HETE-2}/FREGATE. The hypothesis is marginally acceptable in the 25-50 keV and 350-1000 keV energy bands, and confidently accepted in the 50-100 and 100-350 keV bands. Noticing that BATSE is sensitive in 50-350 keV, we conclude that the BATSE observation is consistent with GBM observation in the 50-350 keV band, similar to that shown in Figure \ref{T90_HR}.

\begin{figure*}
\includegraphics[angle=0,scale=0.350,width=0.5\textwidth,height=0.3\textheight]{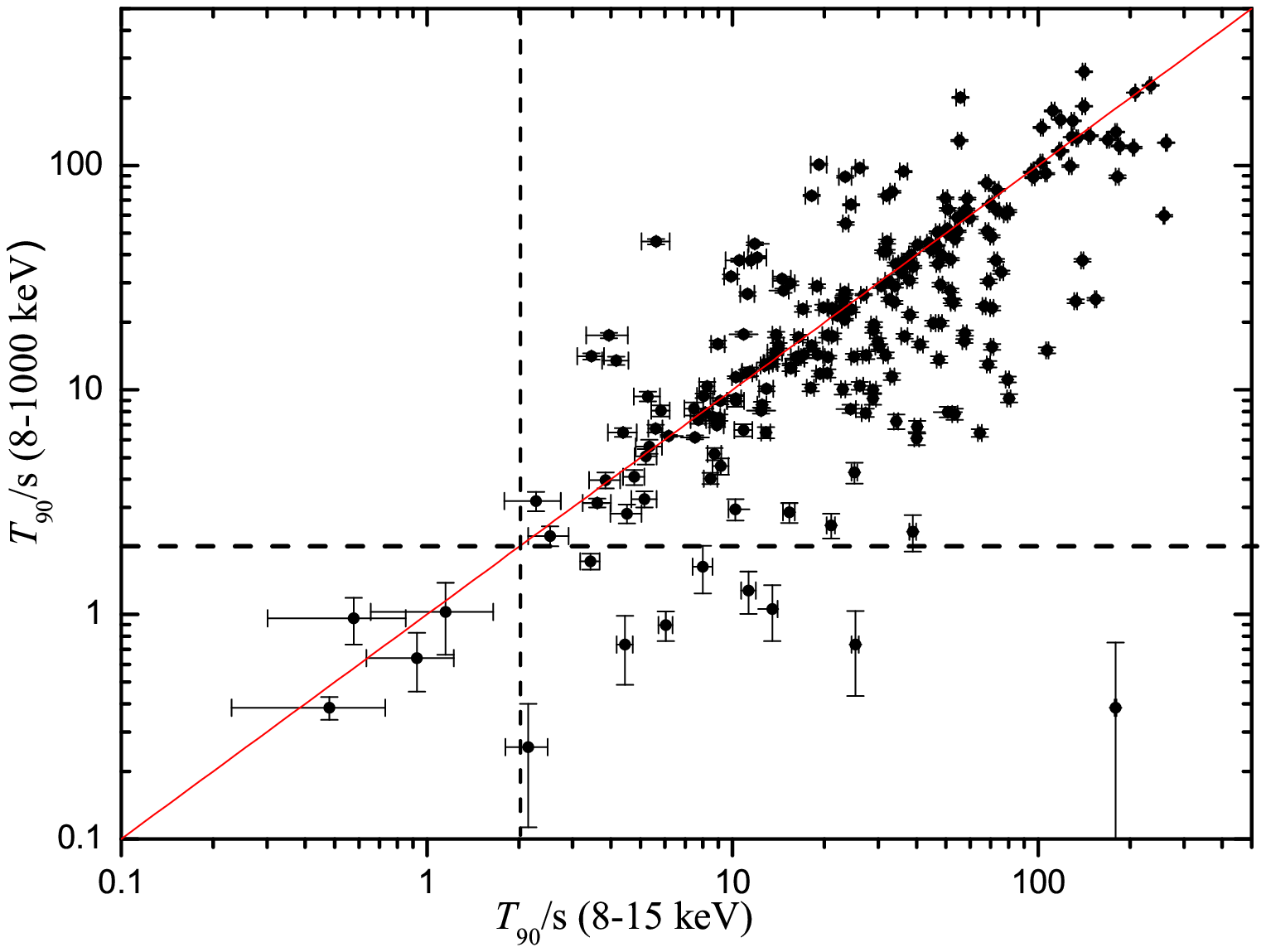}
\includegraphics[angle=0,scale=0.350,width=0.5\textwidth,height=0.3\textheight]{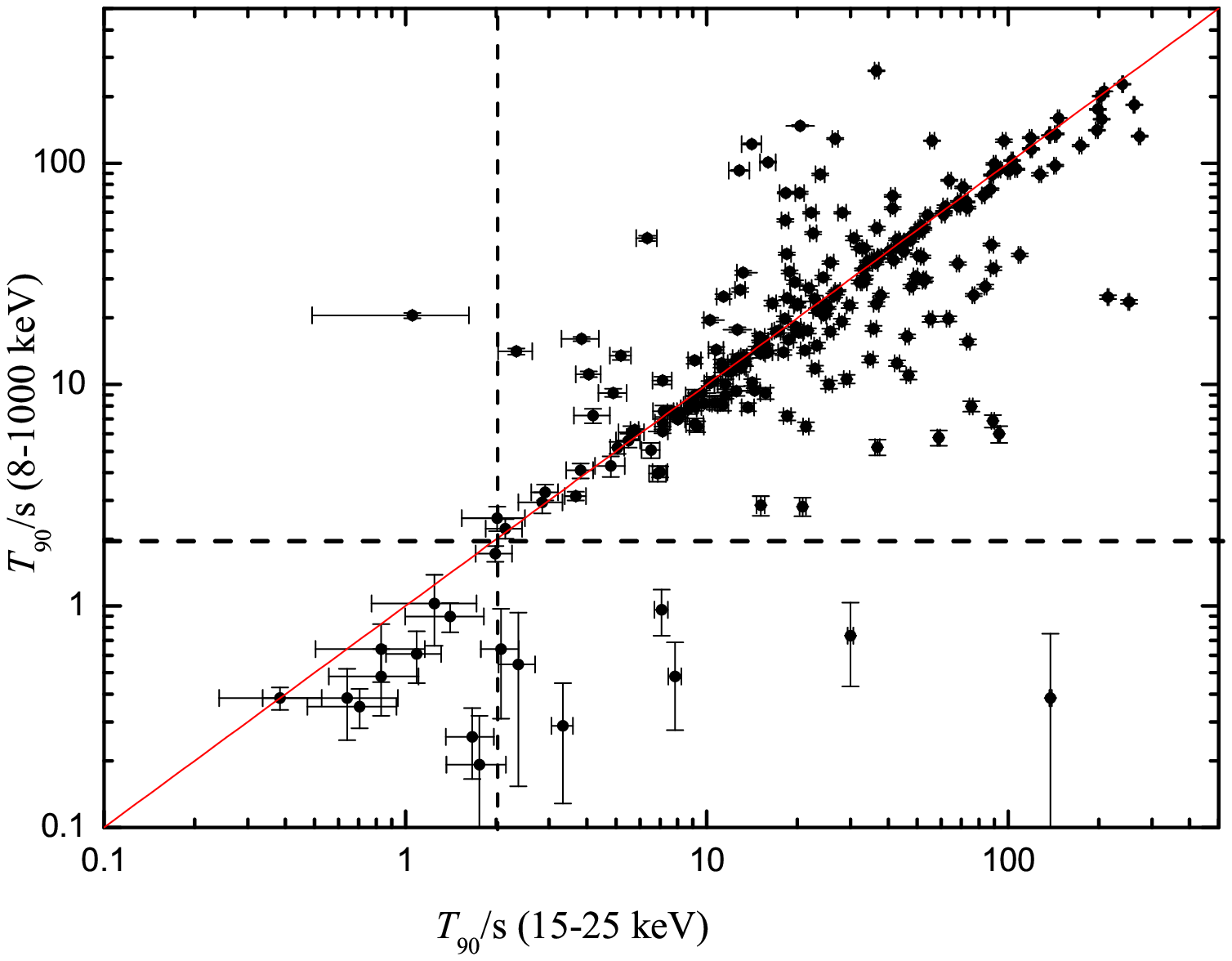}
\includegraphics[angle=0,scale=0.350,width=0.5\textwidth,height=0.3\textheight]{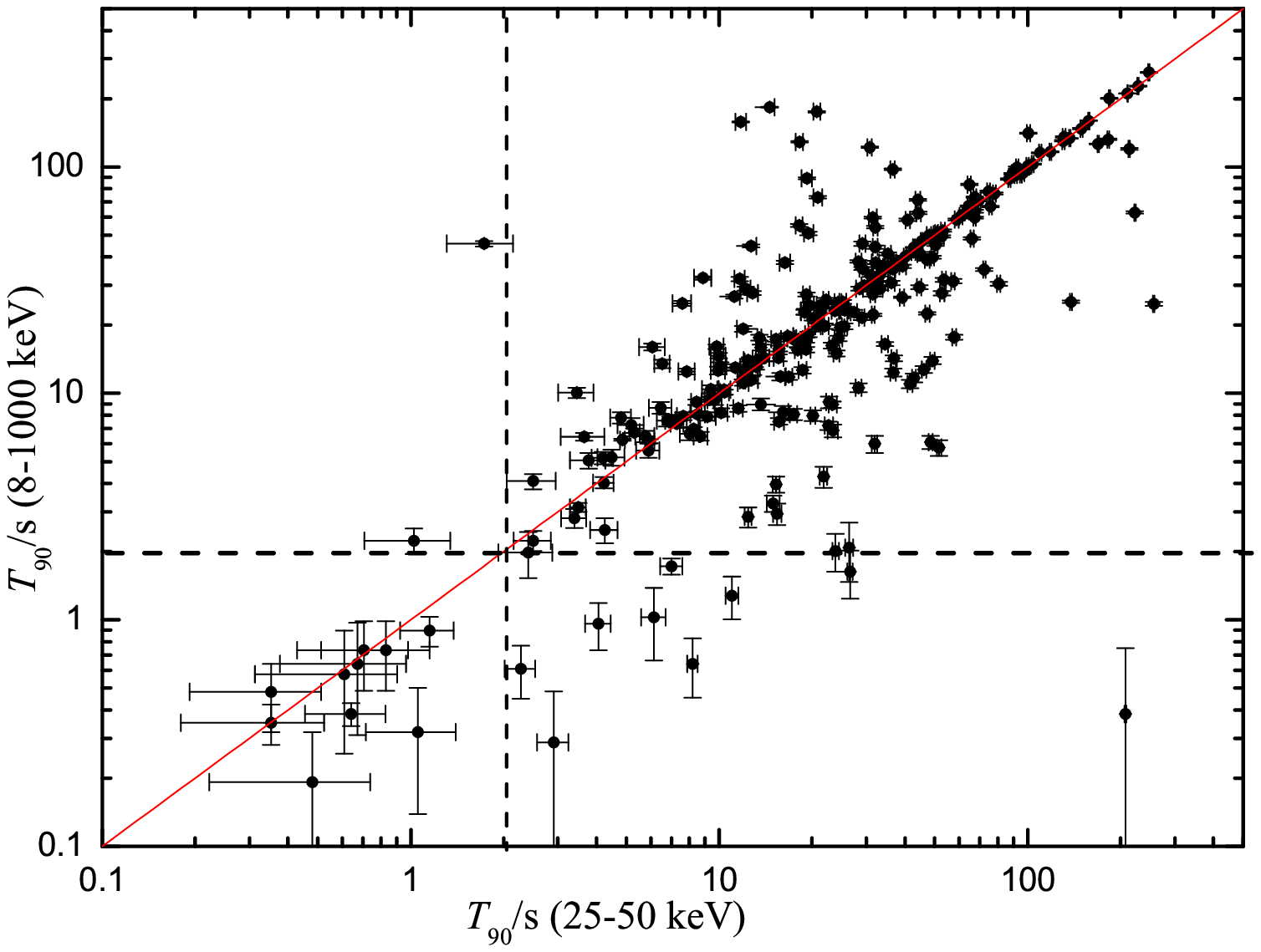}
\includegraphics[angle=0,scale=0.350,width=0.5\textwidth,height=0.3\textheight]{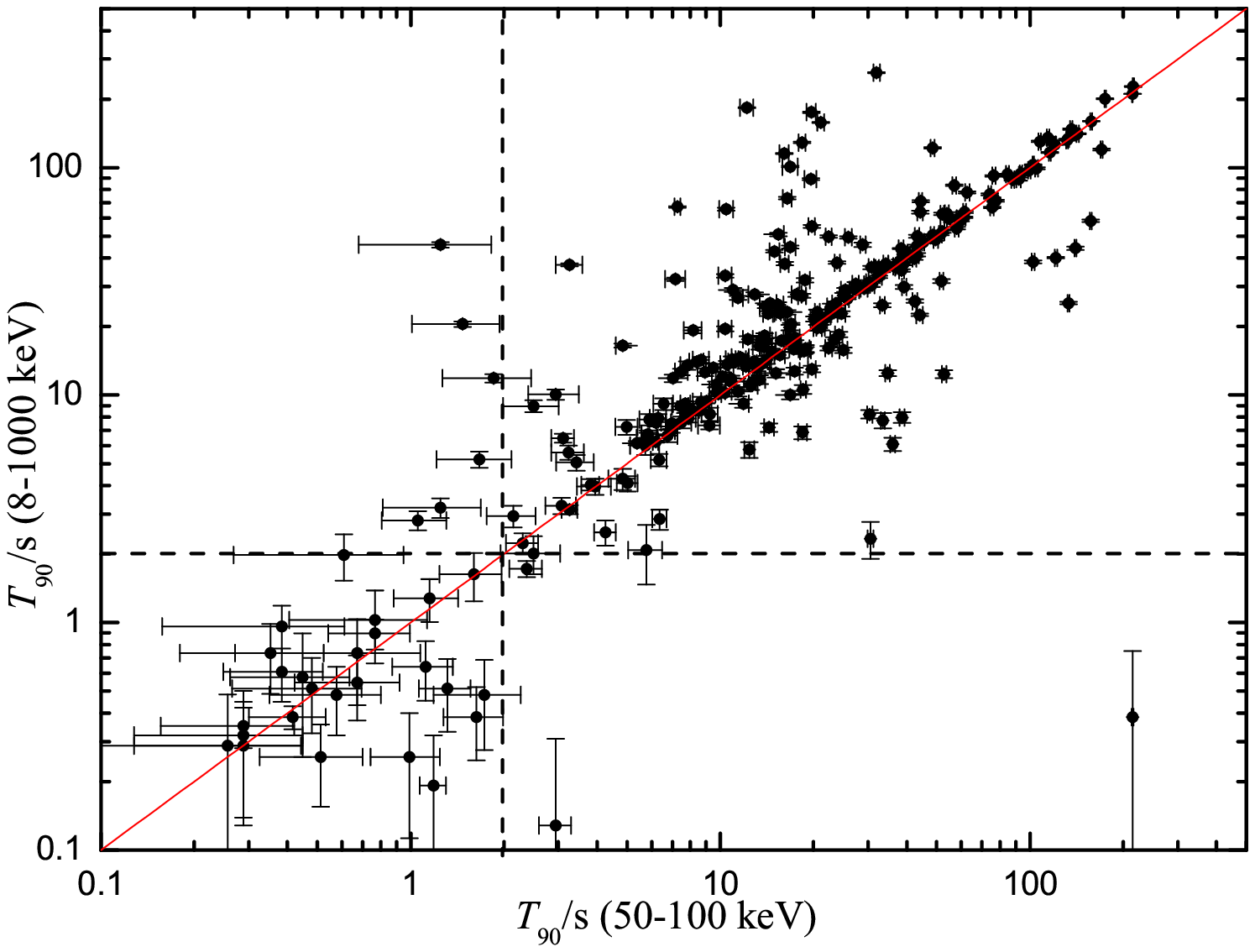}
\includegraphics[angle=0,scale=0.350,width=0.5\textwidth,height=0.3\textheight]{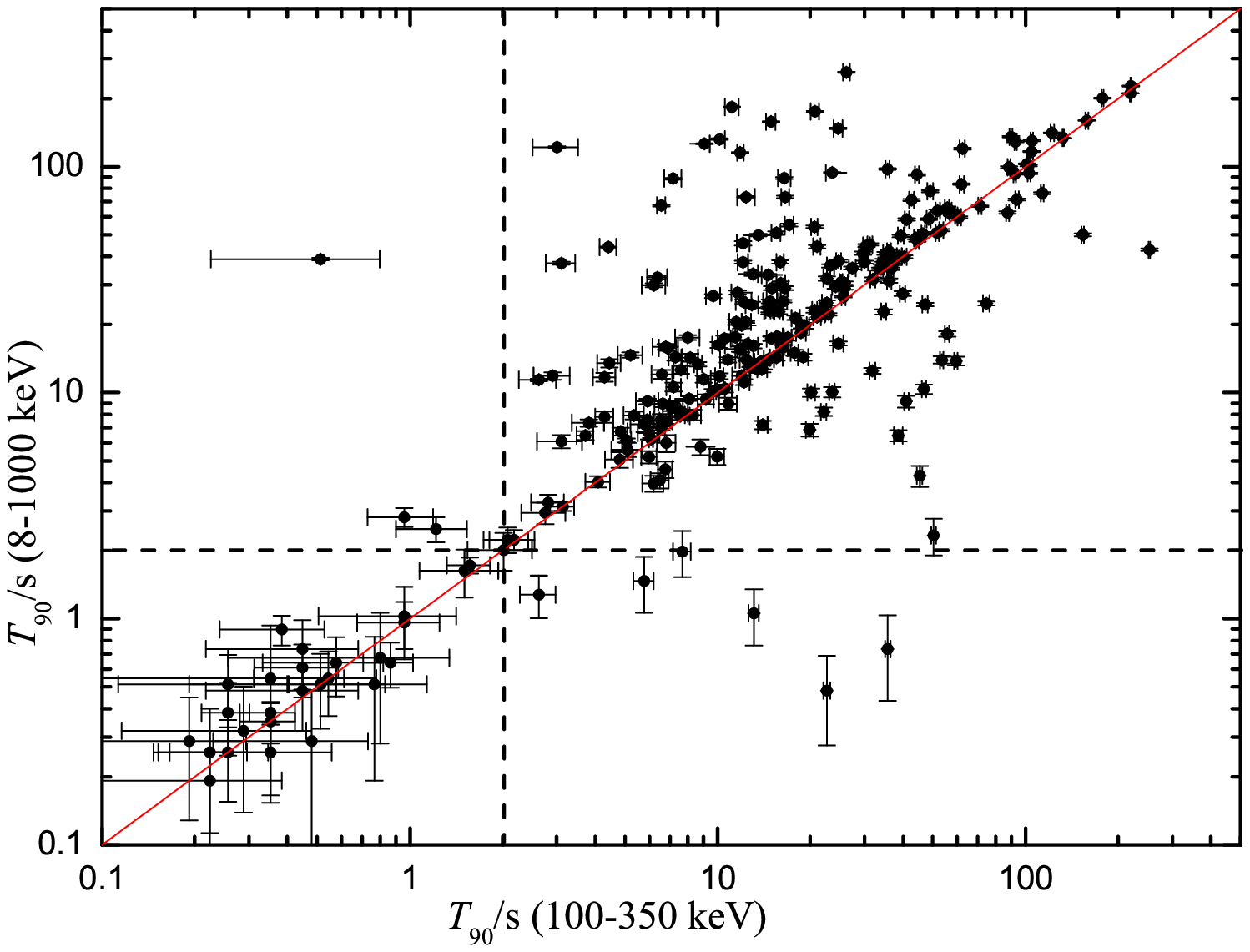}
\includegraphics[angle=0,scale=0.350,width=0.5\textwidth,height=0.3\textheight]{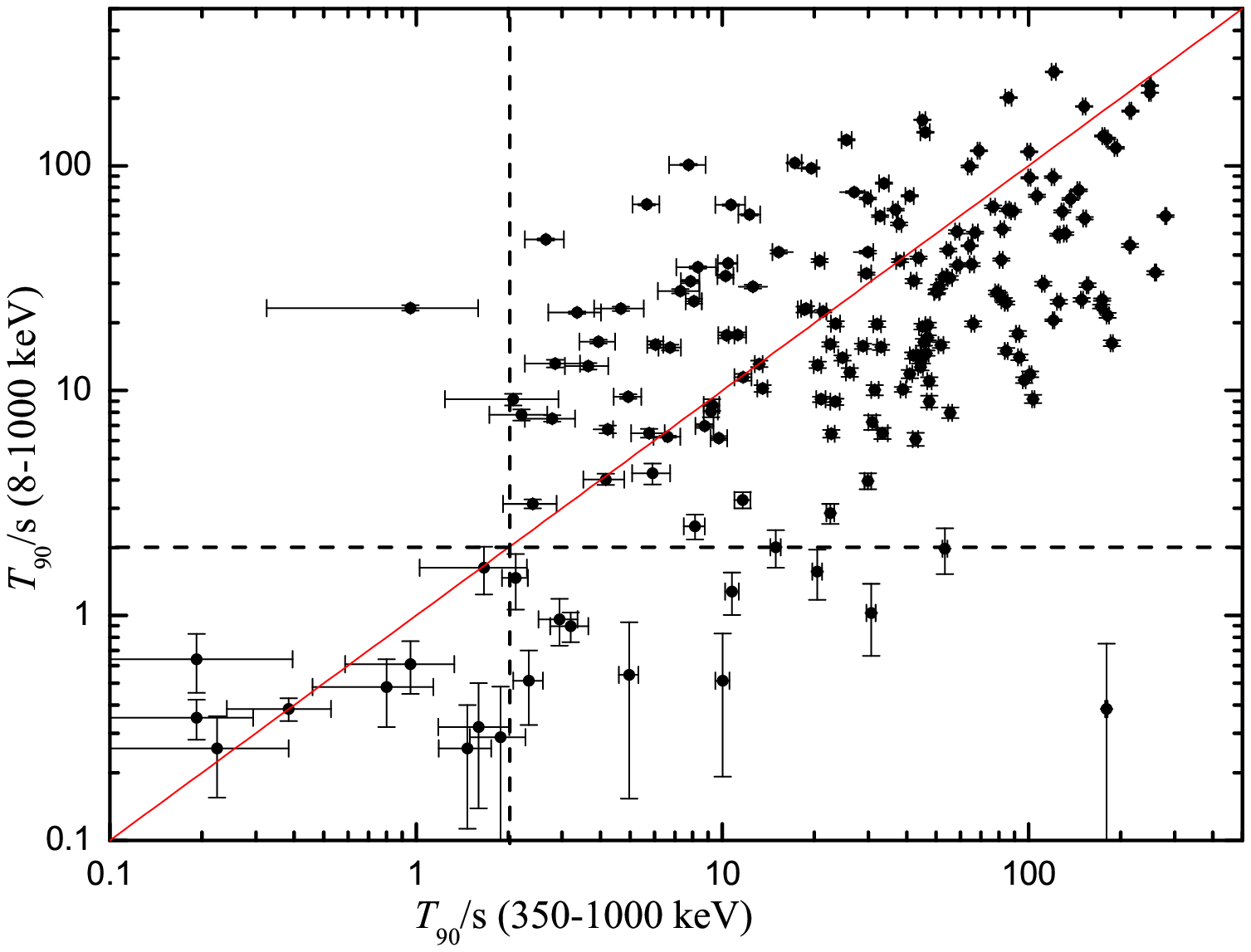}
\caption{Comparisons between $T_{90}$ measured in the 8-1000 keV energy band and some sub-energy-band. The dotted line denotes $T_{90}=2$ seconds and the solid lines are the equality line. }\label{T90_correlation}
\end{figure*}

\begin{figure*}
\centering
\includegraphics[angle=0,scale=0.35]{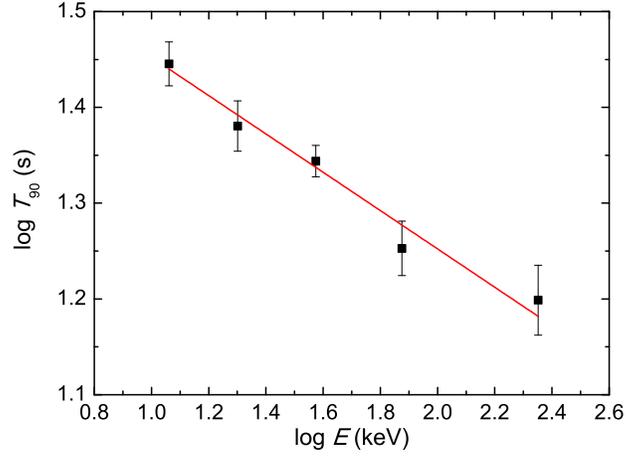}
\caption{Energy dependence of $T_{90}$ for the LGBRs in our sample. The solid line is the best fit to the data.}\label{T90_Energy}
\end{figure*}

We compare the $T_{90}$ in the 8-1000 keV band with that in the sub-energy bands in Figure \ref{T90_correlation}. We still adopt $T_{90}=2$ seconds in the 8-1000 keV band as the division line to classify LGRBs and SGRBs. One can find that some SGRBs in the 8-1000 keV band move to the LGRB group in softer bands. We investigate energy dependence of $T_{90}$ for LGRBs only, since the SGRB sample is too small to give robust statistical result. The typical value of $\bar{T}_{90}$ and its error $\Delta \log T_{90}$ for a given energy band are derived from a Gaussian fit to the $\log T_{90}$ distribution. Figure \ref{T90_Energy} shows $\bar{T}_{90}$ as a function of the central value of the energy band. A clear correlation is found, and the best linear fit gives $\bar{T}_{90}\propto E^{-0.20\pm 0.02}$. Note that the slopes are shallower than those observed in bright GRBs as reported by Richardson et al. (1996) and Bissaldi et al. (2011), who found $\bar{T}_{90}\propto E^{-0.4}$. This may be caused by a sample selection effect. A power-law index $\sim 0.2\sim 0.3$ was also reported in the literature for energy dependence of GRB durations and pulse durations in some GRBs (e.g., Fenimore et al. 1995; Norris et al. 2005; Liang et al. 2006; Zhang 2008).

\section{Extended Central Engine Activity Time in the X-ray and Optical Bands}
As shown above, the burst duration is energy dependence and the bimodal $T_{90}$ distribution would be due to instrumental selection effect. This suggests that if one goes to even lower energy bands, the duration can be even longer. Robotic optical telescopes also detected significant optical flares in some GRBs. Li et al. (2012) presented a detail analysis on the optical flares. They got twenty-four optical flares from 19 GRBs and found that the isotropic flare peak luminosity ($L_{\rm R, iso}$) is correlated with that of gamma-rays, i.e., $L_{\rm R, iso}\propto L_{{\gamma}, \rm iso}^{1.11\pm 0.27}$. The flares peak at from tens of seconds to several days post the GRB trigger. Later flares tend to be wider and dimmer, following the relations  $w\sim t_{\rm p}/2$ and $L_{\rm R, iso}\propto  [t_{\rm p}/(1+z)]^{-1.15\pm0.15}$.  These results suggest that the optical flares are also related to the erratic behavior of the central engine.

\begin{figure*}
\centering
\includegraphics[angle=0,scale=0.35,width=0.25\textwidth,height=0.25\textheight]{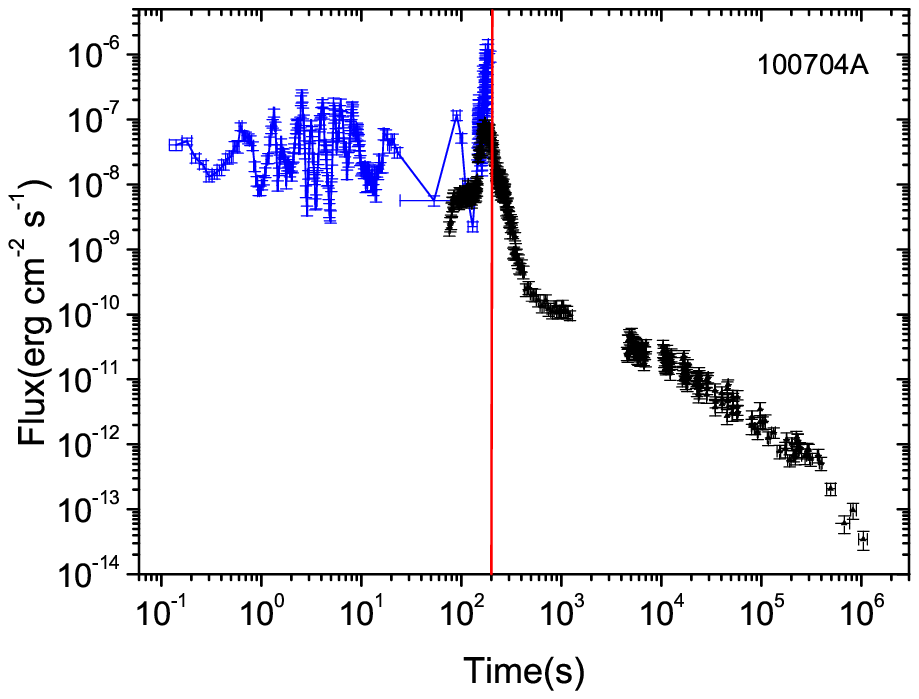}
\includegraphics[angle=0,scale=0.35,width=0.25\textwidth,height=0.25\textheight]{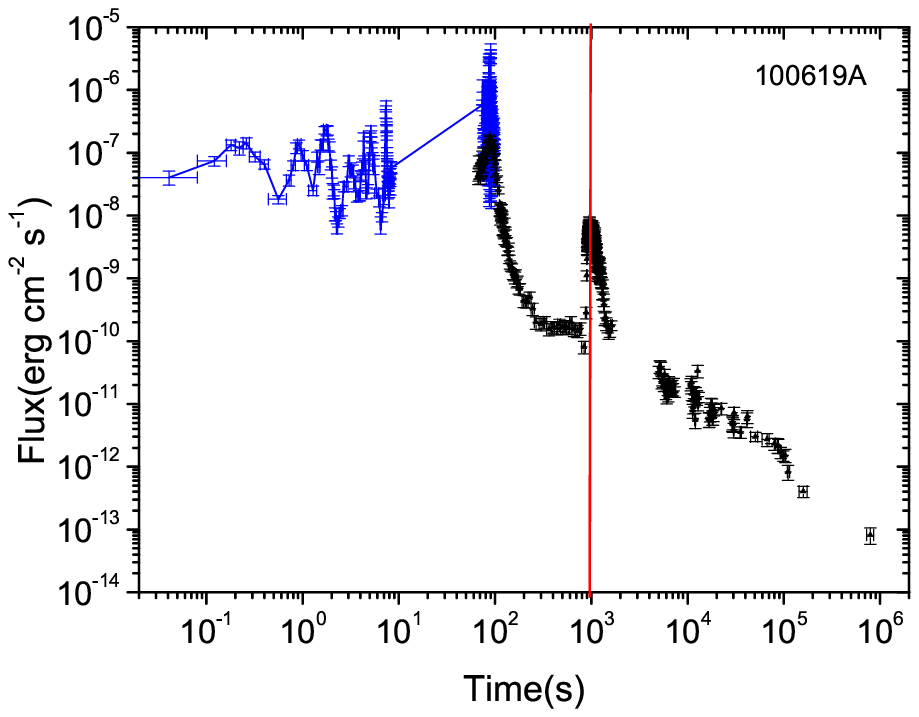}
\includegraphics[angle=0,scale=0.35,width=0.25\textwidth,height=0.25\textheight]{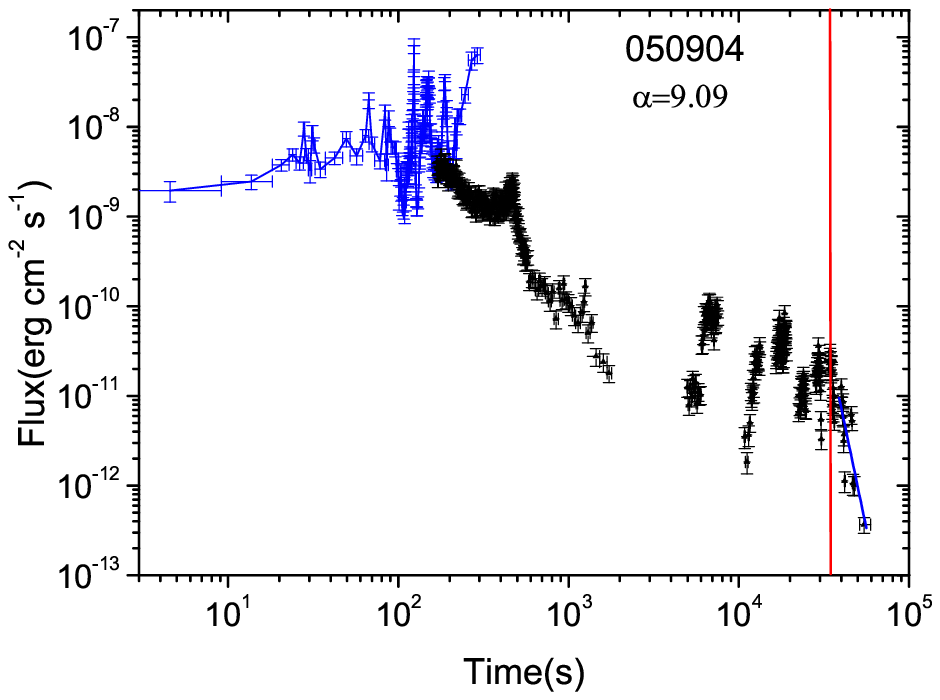}
\caption{Examples of XRT light curves without ({\em left panel}) or with ({\em middle and right panels}) significant flares after $T_{90}$.
The connected data points are the extrapolation of the BAT data to the XRT energy band. Vertical lines correspond to the peak of last X-ray flare, which is used to define the total central engine time scale $T_{\rm f}$.}\label{flares}
\end{figure*}

\begin{figure*}
\centering
\includegraphics[angle=0,scale=0.5,width=0.5\textwidth,height=0.35\textheight]{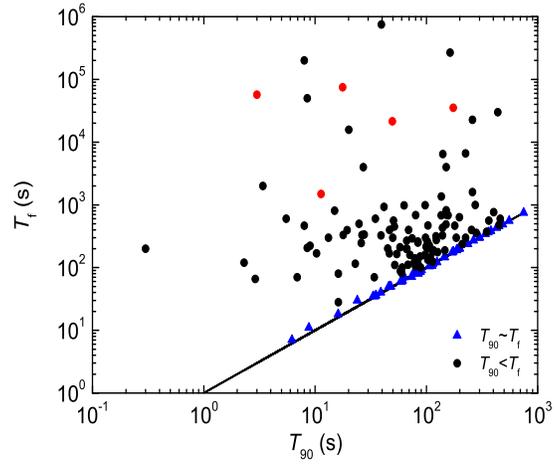}
\caption{The total central engine time scale ($T_{\rm f}$) that is defined with the peak time of the last X-ray flare as a function of $T_{90}$ for our {\em Swift} GRB sample. The triangles denote the GRBs whose $T_{90}\sim T_{\rm f}$ and the solid dots are for those GRBs with $T_{\rm f}>T_{90}$. The solid red dots are the GRBs that their X-ray emission is dominated by flares. The line is the equality line.}\label{T90_Tf}
\end{figure*}

The rapid slewing capacity of the X-ray telescope (XRT) onboard {\em Swift} makes it possible to catch the X-ray emission from very early to late episodes of GRBs. Erratic flares were detected for both LGRBs and SGRB (Burrows et al. 2005; Chincarini et al. 2007; Margutti et al. 2010). The detection probability of X-ray flares is much larger than that of optical flares (Li et al. 2012). It is generally believed that these X-ray flares are due to extended central engine activity at late times (Burrows et al. 2005; King et al. 2005; Fan \& Wei 2005; Zhang et al. 2006; Dai et al. 2006; Perna et al. 2006; Proga \& Zhang 2006; Liang et al. 2006; Lazzati \& Perna 2007; Maxham \& Zhang 2009). We take the peak time of the last X-ray flare to define the central engine duration of a burst (denoted as $T_{\rm f}$). We make an extensive search for the significant flares in the {\em Swift}/XRT lightcurves with the following criteria. First, there are significant flares in the X-ray lightcurve, with $\Delta$F/F $\geq 5$, where $\Delta F=F_{\rm p}-F$ is the flux over the underlying flux level $F$. Second, BAT and XRT lightcurves are well connected without a gap, or significant flares are observed after a gap. We show some examples of these lightcurves in Figure \ref{flares}. We obtain a sample of 159 GRBs. Figure \ref{T90_Tf} shows $T_{\rm f}$ as a function of $T_{90}$. We find that there is no significant flare after $T_{90}$ in 49 GRBs. This suggests that the $T_{90}$s of these bursts are comparable to the durations of the central engines. On the other hand, significant flares are observed in 110 GRBs after their $T_{90}$, indicating the $T_{\rm f}$ of these bursts are much larger than $T_{90}$. The X-ray emission of four GRBs, 050502B, 050724, 050904, and 060223, are dominated by flares (as shown in Figure \ref{T90_Tf}), indicating that they may be super-LGRBs (Zou et al. 2006).

\section{ Conclusions and Discussion}
We have calculated $T_{90}$ of {\em Fermi}/GBM GRBs in various energy bands and compared the $T_{90}$ distribution with those obtained from previous/current GRB survey missions. We show that the $T_{90}$ distribution in the 8-1000 keV band is bimodal, being roughly consistent with that of the {\em CGRO}/BATSE GRB sample, but short-to-long GRB number ratio is 1:5, being lower than that in the BATSE sample (1:3). We measure the $T_{90}$ in several sub-bands, i.e., 8-15, 15-25, 25-50, 50-100, 100-350, and 350-1000 keV bands to investigate the energy band selection effect on the bimodal behavior of $T_{90}$ distribution. It is found that the bimodal feature is well recognized in the 50-100 and 100-350 keV bands and only marginally accepted in the 25-50 keV and 350-1000 keV energy bands. The hypothesis of the bimodality is confidently rejected in 8-15 and 15-25 keV bands. We compare the $T_{90}$ distributions in these sub-energy bands with those derived from other GRB detectors with similar energy bands and find that they are roughly consistent with each other. $T_{90}$ as a function of energy band follows ${\bar T}_{90}>\propto E^{-0.20\pm 0.02}$ for LGRBs. Some GRBs fall into the short category in a high energy band, but move to the long category in a lower energy band. Considering the erratic optical and X-ray flares that may have the same physical origin as the prompt gamma-rays, the duration of a burst would be even much longer for most GRBs. These results indicate that $T_{90}$ is energy dependent and the bimodal $T_{90}$ distribution is valid only for certain energy bands.

Burst duration is critical for both GRB classification and understanding the behavior of GRB central engine. It is an indicator of the lifetime of GRB central engine. Popular central engine models of GRBs are related to accretion onto a central black hole that is formed from collapse of a massive star or merger of a compact star binary. Current favored jet launching models for GRBs include neutrino-annihilation mechanism from a neutrino-dominated accretion flow (NDAFs, e.g., Popham et al. 1999; Narayan et al. 2001; Kohri \& Mineshige 2002; Di Matteo et al. 2002; Kohri et al. 2005; Gu et al. 2006; Chen \& Beloborodov 2007; Liu et al. 2007; 2010; Xie et al. 2007; Lei et al. 2009) and Blandford-Znajek process (Blandford-Znajek 1977; Lee et al. 2000; Lei et al. 2007). It is theoretically expected that the accreting time scale for compact star mergers would be shorter than 1 $\rm s$  based on both analytical and numerical results (e.g., Narayan et al. 2001; Aloy et al. 2005). The $T_{90}$ distribution observed with {\em CGRO}/BATSE (Kouveliotou et al. 1993) seems to be consistent with the speculation that two types of GRBs (long vs. short) are consistent with two distinct progenitors, i.e. collapses of massive stars vs. mergers of compact objects. However, as we have shown here, the bimodal $T_{90}$ distribution would be likely due to an instrumental selection effect. Our results, together with the observed extended emission of some short GRBs, not only challenge the long-short GRB classification scheme, but also challenge the conventional GRB central engine models, and call for new mechanisms to account for extended GRB central engine activities (e.g., King et al. 2005; Fan \& Wei 2005; Zhang et al. 2006; Dai et al. 2006; Perna et al. 2006; Proga \& Zhang 2006; Metzger et al. 2008; Liu et al. 2012).

\acknowledgments We acknowledge the use of the public data from the Swift data archive. We appreciate helpful comments from the referees. This work is supported by the ``973" Program of China (2009CB824800), the National Natural Science Foundation of China (Grants No. 11025313, 11203008, 11078008, 11063001, and 11163001), the Special Foundation for Distinguished Expert Program of Guangxi, the Guangxi Natural Science Foundation
(2010GXNSFA013112, 2011GXNSFB018063 and 2010GXNSFC013011), the special funding for national outstanding young scientist (Contract No. 2011-135), and the 3th Innovation Projet of Guangxi University. BZ acknowledges support from NSF (AST-0908362).

\begin{deluxetable}{ccccccccc}
\tabletypesize{\scriptsize}
\tablecaption{Derived GBM $T_{90}$ in different energy bands$^{*}$}
\tablewidth{0pt}
\tabletypesize{\tiny}
\tablehead{
\colhead{GBM ID} &
\colhead{GRB} &
\colhead{8-15 keV} &
\colhead{15-25 keV} &
\colhead{25-50 keV} &
\colhead{50-100 keV} &
\colhead{100-350 keV} &
\colhead{350-1000 keV} &
\colhead{8-1000 keV} \\
\colhead{} &
\colhead{name} &
\colhead{(s)} &
\colhead{(s)} &
\colhead{(s)} &
\colhead{(s)} &
\colhead{(s)} &
\colhead{(s)}&
\colhead{(s)}
}
\startdata
080714745&080714&--&18.85$\pm$0.62&8.86$\pm$0.57&7.17$\pm$0.54&6.37$\pm$0.48&10.27$\pm$0.62&32.29$\pm$0.54\\
080725435&080725&21.79$\pm$0.58&25.12$\pm$0.44&31.58$\pm$0.4&22.27$\pm$0.46&22.88$\pm$0.54&3.36$\pm$0.66&22.21$\pm$0.23\\
080727964&080727C&--&25.7$\pm$0.66&29.12$\pm$0.66&32.77$\pm$0.66&27.36$\pm$0.69&--&35.55$\pm$0.59\\
080804972&080804&31.87$\pm$0.72&20.35$\pm$0.66&20.86$\pm$0.7&16.45$\pm$0.53&16.58$\pm$0.46&106.5$\pm$0.98&73.41$\pm$0.93\\
080810549&080810&50.91$\pm$0.99&51.39$\pm$0.76&52.7$\pm$0.83&25.92$\pm$0.69&39.39$\pm$0.88&125.28$\pm$1.12&49.34$\pm$0.63\\
080904886&080904&41.18$\pm$0.88&18.72$\pm$0.7&13.66$\pm$0.49&17.28$\pm$0.6&6.78$\pm$0.65&51.87$\pm$0.95&15.94$\pm$0.47\\
080905499&080905A&13.5$\pm$0.57&--&--&--&13.09$\pm$0.5&--&1.06$\pm$0.3\\
080905570&080905C&23.26$\pm$0.72&21.79$\pm$0.53&19.2$\pm$0.51&18.37$\pm$0.51&--&78.82$\pm$1&27.36$\pm$0.56\\
080905705&080905B&129.73$\pm$1.48&204.64$\pm$1.16&11.74$\pm$0.48&21.12$\pm$0.57&14.88$\pm$0.53&--&158.02$\pm$1.10\\
080906212&080906B&5.15$\pm$0.49&2.91$\pm$0.3&14.98$\pm$0.73&3.07$\pm$0.34&2.82$\pm$0.34&11.65$\pm$0.7&3.26$\pm$0.26\\
080912360&080912&57.63$\pm$0.83&35.78$\pm$0.75&16.67$\pm$0.72&17.44$\pm$0.62&--&92.29$\pm$1.16&17.89$\pm$0.49\\
080913735&080913B&32.64$\pm$0.78&37.73$\pm$0.5&25.18$\pm$0.62&14.5$\pm$0.67&14.82$\pm$0.76&149.54$\pm$1.48&25.31$\pm$0.52\\
080916009&080916C&58.24$\pm$0.85&68.54$\pm$0.95&63.52$\pm$1.04&61.47$\pm$0.62&56.96$\pm$0.58&36.77$\pm$0.74&63.65$\pm$0.61\\
080916406&080916A&40.35$\pm$0.7&44.58$\pm$0.7&32.03$\pm$0.6&140.22$\pm$1.14&20.99$\pm$0.57&214.88$\pm$1.39&44.26$\pm$0.72\\
080920268&080920&140.42$\pm$1.54&--&--&71.94$\pm$1.2&--&73.09$\pm$1.38&--\\
080925775&080925&19.81$\pm$0.62&16.51$\pm$0.47&25.5$\pm$0.76&15.58$\pm$0.44&14.72$\pm$0.4&0.96$\pm$0.64&23.33$\pm$0.62\\
080927480&080927&70.62$\pm$0.9&36.51$\pm$0.72&23.74$\pm$0.65&16.45$\pm$0.75&15.39$\pm$0.53&4.67$\pm$0.86&23.14$\pm$0.54\\
080928628&080928&33.86$\pm$0.51&18.5$\pm$0.49&21.44$\pm$0.69&14.05$\pm$0.66&47.2$\pm$0.99&84.48$\pm$0.91&24.54$\pm$0.44\\
081006604&081006&9.15$\pm$0.55&--&--&--&6.75$\pm$0.36&--&4.58$\pm$0.39\\
081006872&081006B&3.46$\pm$0.77&34.66$\pm$0.73&9.79$\pm$0.34&--&6.5$\pm$0.39&--&--\\
081008832&081008&111.94$\pm$1.39&198.66$\pm$1.2&20.7$\pm$0.52&19.68$\pm$0.68&20.7$\pm$0.62&215.3$\pm$1.54&175.2$\pm$1.16\\
081009140&081009&46.59$\pm$0.33&45.47$\pm$0.36&42.78$\pm$0.3&38.37$\pm$0.74&4.42$\pm$0.28&64.22$\pm$0.73&44.16$\pm$0.2\\
081012549&081012&25.02$\pm$0.59&--&--&12.86$\pm$0.6&14.66$\pm$0.55&93.44$\pm$1.22&14.08$\pm$0.42\\
081021398&081021&--&--&--&15.87$\pm$0.36&11.81$\pm$0.36&--&--\\
081024245&0081024&--&--&--&2.94$\pm$0.34&--&--&0.13$\pm$0.18\\
081024891&081024B&--&--&107.23$\pm$1.05&1.25$\pm$1.08&50.72$\pm$0.94&86.4$\pm$0.9&--\\
081025349&081025&66.59$\pm$0.92&251.97$\pm$1.42&25.38$\pm$0.59&23.74$\pm$0.46&21.95$\pm$0.47&173.09$\pm$1.71&23.62$\pm$0.49\\
081028538&081028B&12.86$\pm$0.46&9.28$\pm$0.52&5.76$\pm$0.41&5.82$\pm$0.45&38.62$\pm$1.24&33.41$\pm$1.27&6.46$\pm$0.36\\
081101491&081101&--&2.37$\pm$0.33&--&--&0.35$\pm$0.26&4.96$\pm$0.36&0.54$\pm$0.39\\
081101532&081101B&12.42$\pm$0.55&8.86$\pm$0.49&8.51$\pm$0.46&7.71$\pm$0.3&7.78$\pm$0.2&--&8.1$\pm$0.25\\
081102365&081102B&38.94$\pm$0.96&--&--&30.59$\pm$0.66&50.34$\pm$0.9&--&2.34$\pm$0.43\\
081102739&081102&47.94$\pm$0.77&52.54$\pm$0.65&44.54$\pm$0.72&29.12$\pm$0.72&24.16$\pm$0.63&156.06$\pm$1.23&29.47$\pm$0.59\\
081105614&081105B&14.91$\pm$0.62&--&12.77$\pm$0.44&--&8.03$\pm$0.48&14.82$\pm$0.63&--\\
081107321&081107&3.42$\pm$0.25&1.98$\pm$0.28&7.01$\pm$0.58&2.37$\pm$0.29&1.57$\pm$0.25&--&1.73$\pm$0.14\\
081109293&081109A&--&--&31.46$\pm$0.54&24.96$\pm$0.66&39.87$\pm$0.77&50.11$\pm$0.85&27.46$\pm$0.65\\
081110601&081110&19.3$\pm$0.54&22.91$\pm$0.63&16.8$\pm$0.46&10.85$\pm$0.46&10.14$\pm$0.32&101.41$\pm$0.92&11.81$\pm$0.36\\
081113230&081113&4.45$\pm$0.27&--&0.83$\pm$0.32&--&0.45$\pm$0.23&--&0.74$\pm$0.25\\
081118876&081118&20.42$\pm$0.72&17.34$\pm$0.57&17.22$\pm$0.57&13.54$\pm$0.59&8$\pm$0.74&--&17.54$\pm$0.47\\
081119184&081119&4.16$\pm$0.4&3.87$\pm$0.32&--&0.93$\pm$0.3&--&--&--\\
081120618&081120&28.99$\pm$0.62&10.24$\pm$0.47&24.74$\pm$0.69&10.37$\pm$0.52&--&46.91$\pm$0.93&19.52$\pm$0.49\\
081121858&081121&23.46$\pm$0.52&18.21$\pm$0.48&18.18$\pm$0.46&19.74$\pm$0.49&17.09$\pm$0.57&37.82$\pm$0.73&55.20$\pm$0.84\\
081122520&081122&3.94$\pm$0.61&21.63$\pm$0.62&24.38$\pm$0.72&23.26$\pm$0.62&16.9$\pm$0.52&--&17.5$\pm$0.48\\
081122614&081122B&--&--&1.02$\pm$0.32&--&2.08$\pm$0.35&--&2.24$\pm$0.29\\
081124060&081124&22.78$\pm$0.75&19.97$\pm$0.51&18.88$\pm$0.47&14.27$\pm$0.51&15.78$\pm$0.52&--&22.82$\pm$0.58\\
081125496&081125&9.12$\pm$0.48&11.55$\pm$0.53&23.33$\pm$0.72&7.42$\pm$0.45&6.62$\pm$0.39&23.36$\pm$0.78&8.93$\pm$0.3\\
081126899&081126&39.1$\pm$0.98&36.26$\pm$0.69&36.54$\pm$0.73&38.5$\pm$0.66&36.54$\pm$0.75&8.32$\pm$1.23&35.36$\pm$0.46\\
081129161&081129&9.89$\pm$0.48&13.18$\pm$0.6&11.65$\pm$0.43&18.75$\pm$0.71&22.62$\pm$0.6&53.02$\pm$0.87&32$\pm$0.75\\
081130212&081130&--&--&44.35$\pm$0.69&--&--&--&--\\
081130629&081130B&4.16$\pm$0.41&5.18$\pm$0.41&6.53$\pm$0.36&7.81$\pm$0.36&4.45$\pm$0.39&--&13.5$\pm$0.59\\
081204004&081204C&10.21$\pm$0.49&2.85$\pm$0.48&15.42$\pm$0.47&2.14$\pm$0.39&2.75$\pm$0.45&--&2.94$\pm$0.32\\
081204517&081204B&--&0.83$\pm$0.28&0.35$\pm$0.16&0.58$\pm$0.23&0.45$\pm$0.23&0.8$\pm$0.34&0.48$\pm$0.16\\
081206275&081206&--&55.26$\pm$0.84&25.47$\pm$0.53&20.61$\pm$0.66&19.2$\pm$0.65&32.06$\pm$0.95&19.74$\pm$0.59\\
081206604&081206B&80.29$\pm$1.05&4.9$\pm$0.52&--&11.87$\pm$0.49&--&103.58$\pm$0.86&9.18$\pm$0.39\\
081206987&081206C&134.75$\pm$1.11&48.03$\pm$0.64&--&--&126.02$\pm$1.2&--&--\\
081207680&081207&127.58$\pm$1.27&90.11$\pm$0.77&92.42$\pm$0.85&104.99$\pm$1.02&88.48$\pm$1.02&64.26$\pm$0.99&99.39$\pm$0.85\\
081209981&081209&--&--&--&--&0.45$\pm$0.32&--&--\\
081213173&081213&1.73$\pm$0.37&--&--&--&--&3.14$\pm$0.29&--\\
081215784&081215&5.6$\pm$0.3&7.14$\pm$0.42&5.31$\pm$0.2&5.82$\pm$0.26&4.83$\pm$0.26&4.22$\pm$0.18&6.72$\pm$0.27\\
081215880&081215B&--&--&18.66$\pm$0.56&7.49$\pm$0.36&13.5$\pm$0.6&--&12.67$\pm$0.42\\
081216531&081216&6.05$\pm$0.32&1.41$\pm$0.41&1.15$\pm$0.23&0.77$\pm$0.23&0.38$\pm$0.14&3.2$\pm$0.45&0.9$\pm$0.14\\
081217983&081217&23.33$\pm$0.58&24.45$\pm$0.42&20.67$\pm$0.41&16.96$\pm$0.52&12.38$\pm$0.44&120.64$\pm$0.97&20.54$\pm$0.23\\
081221681&081221&32$\pm$0.97&30.75$\pm$0.72&29.15$\pm$0.63&28.86$\pm$0.62&12.1$\pm$0.57&--&45.82$\pm$0.99\\
081222204&081222&11.2$\pm$0.6&12.9$\pm$0.5&11.2$\pm$0.47&11.39$\pm$0.43&9.66$\pm$0.52&80.1$\pm$1.16&26.75$\pm$0.53\\
081223419&081223&--&1.09$\pm$0.23&2.27$\pm$0.26&0.38$\pm$0.14&0.45$\pm$0.14&0.96$\pm$0.37&0.61$\pm$0.16\\
081224887&081224&8.06$\pm$0.41&14.4$\pm$0.59&9.28$\pm$0.33&9.02$\pm$0.34&8.1$\pm$0.36&4.93$\pm$0.51&9.41$\pm$0.2\\
081226044&081226&--&7.84$\pm$0.39&--&1.73$\pm$0.54&22.62$\pm$0.6&--&0.48$\pm$0.2\\
081226156&081226C&16.51$\pm$0.69&15.04$\pm$0.62&12.61$\pm$0.51&13.76$\pm$0.66&59.42$\pm$0.78&44.83$\pm$1.16&13.79$\pm$0.6\\
081226509&081226B&--&--&--&--&0.77$\pm$0.36&10.02$\pm$0.53&0.51$\pm$0.32\\
081229187&081229&--&--&--&--&0.86$\pm$0.16&--&0.64$\pm$0.14\\
081231140&081231&26.91$\pm$0.59&27.17$\pm$0.52&39.14$\pm$0.63&25.7$\pm$0.34&25.63$\pm$0.36&--&26.56$\pm$0.23\\
090102122&090102&30.72$\pm$0.79&33.18$\pm$0.82&28.51$\pm$0.7&25.31$\pm$0.82&16.51$\pm$0.47&12.61$\pm$0.55&29.02$\pm$0.54\\
090107681&090107B&3.46$\pm$0.34&2.34$\pm$0.3&--&--&--&--&14.11$\pm$0.44\\
090108020&090108&1.15$\pm$0.5&1.25$\pm$0.48&6.14$\pm$0.56&0.77$\pm$0.36&0.96$\pm$0.45&30.62$\pm$1.07&1.02$\pm$0.36\\
090108322&090108B&--&--&--&--&0.8$\pm$0.54&--&0.67$\pm$0.39\\
090109332&090109&15.33$\pm$0.68&15.1$\pm$0.49&12.42$\pm$0.33&6.37$\pm$0.35&--&22.56$\pm$0.68&2.85$\pm$0.30\\
090112332&090112&27.23$\pm$0.72&13.7$\pm$0.65&9.15$\pm$0.6&7.87$\pm$0.45&7.55$\pm$0.56&--&7.9$\pm$0.34\\
090112729&090112B&15.52$\pm$0.64&13.38$\pm$0.57&9.92$\pm$0.52&8.93$\pm$0.48&7.58$\pm$0.39&--&12.61$\pm$0.49\\
090113778&090113&12.48$\pm$0.43&9.02$\pm$0.32&11.52$\pm$0.42&8.67$\pm$0.49&7.07$\pm$0.44&--&8.58$\pm$0.29\\
090117335&090117B&--&--&--&33.5$\pm$0.52&--&--&7.74$\pm$0.59\\
090117640&090117&8.96$\pm$0.46&15.07$\pm$0.5&6.08$\pm$0.57&18.85$\pm$0.58&--&6.05$\pm$0.36&16.03$\pm$0.59\\
090126227&090126B&7.49$\pm$0.54&10.66$\pm$0.44&16.1$\pm$0.68&9.25$\pm$0.57&--&--&8.26$\pm$0.54\\
090126245&090126C&--&--&--&21.44$\pm$0.7&--&23.14$\pm$0.82&--\\
090129880&090129&20.58$\pm$0.53&15.97$\pm$0.53&12.42$\pm$0.43&8.38$\pm$0.45&12.42$\pm$0.63&--&14.02$\pm$0.23\\
090131090&090131&35.36$\pm$0.5&34.24$\pm$0.53&33.47$\pm$0.49&32.06$\pm$0.43&34.75$\pm$0.75&58.91$\pm$0.76&36.06$\pm$0.48\\
090202347&090202&51.65$\pm$0.75&47.81$\pm$0.69&12.83$\pm$0.53&12.9$\pm$0.59&11.62$\pm$0.52&50.37$\pm$0.96&27.74$\pm$0.5\\
090206620&090206&25.22$\pm$0.66&30.05$\pm$0.66&--&0.67$\pm$0.4&35.65$\pm$0.72&--&0.74$\pm$0.3\\
090207777&090207&11.39$\pm$0.64&12.7$\pm$0.66&12.32$\pm$0.57&10.18$\pm$0.51&6.59$\pm$0.51&26.14$\pm$0.77&12.06$\pm$0.52\\
090217206&090217&32.38$\pm$0.82&33.54$\pm$0.65&32.42$\pm$0.56&29.76$\pm$0.65&25.82$\pm$0.68&7.87$\pm$0.55&30.59$\pm$0.53\\
090219074&090219&--&--&--&3.07$\pm$0.28&--&1.44$\pm$0.44&--\\
090222179&090222&30.18$\pm$0.84&73.73$\pm$0.95&19.14$\pm$0.65&18.62$\pm$0.55&11.84$\pm$0.5&32.96$\pm$0.93&15.62$\pm$0.42\\
090227772&090227B&0.58$\pm$0.28&7.07$\pm$0.36&4.06$\pm$0.39&0.38$\pm$0.23&0.96$\pm$0.29&2.94$\pm$0.43&0.96$\pm$0.23\\
090228976&090228B&40.06$\pm$0.75&5.63$\pm$0.54&48.48$\pm$0.74&36$\pm$0.69&3.1$\pm$0.52&42.78$\pm$0.86&6.08$\pm$0.41\\
090301315&090301B&--&93.28$\pm$0.92&31.87$\pm$0.59&5.73$\pm$0.61&6.82$\pm$0.48&--&5.98$\pm$0.52\\
090304216&090304&7.58$\pm$0.89&--&18.18$\pm$0.83&--&--&--&--\\
090305052&090305B&20.99$\pm$0.65&2.02$\pm$0.48&4.26$\pm$0.44&4.26$\pm$0.35&1.22$\pm$0.32&8.13$\pm$0.65&2.5$\pm$0.32\\
090306245&090306C&14.69$\pm$0.62&84.13$\pm$0.9&52.8$\pm$0.91&17.66$\pm$0.64&--&7.3$\pm$1.13&27.71$\pm$0.6\\
090308734&090308B&8$\pm$0.59&--&26.56$\pm$0.53&1.6$\pm$0.36&1.5$\pm$0.43&1.66$\pm$0.64&1.63$\pm$0.39\\
090310189&090310&184.86$\pm$1.33&14.11$\pm$1.06&30.75$\pm$0.53&48.51$\pm$0.67&3.01$\pm$0.51&--&122.11$\pm$0.85\\
090319622&090319&10.53$\pm$1.02&--&32.16$\pm$0.87&34.21$\pm$0.63&37.41$\pm$0.64&38.11$\pm$1.12&37.79$\pm$0.64\\
090320045&090320C&64.54$\pm$0.98&--&3.65$\pm$0.59&--&--&22.72$\pm$0.56&6.43$\pm$0.26\\
090320418&090320A&4.38$\pm$0.48&21.38$\pm$0.47&8.7$\pm$0.45&3.1$\pm$0.26&3.71$\pm$0.23&5.76$\pm$0.73&6.46$\pm$0.29\\
090320801&090320B&15.46$\pm$0.54&52.96$\pm$0.64&29.57$\pm$0.62&39.23$\pm$0.67&6.21$\pm$0.55&111.81$\pm$1.26&29.89$\pm$0.55\\
090323002&090323&78.3$\pm$1.06&61.38$\pm$0.79&60.77$\pm$0.64&61.25$\pm$0.65&59.58$\pm$0.56&12.29$\pm$1&60.51$\pm$0.53\\
090326633&090326&8.26$\pm$0.46&10.56$\pm$0.55&10.37$\pm$0.52&11.39$\pm$0.51&46.82$\pm$0.8&--&10.37$\pm$0.51\\
090327404&090327&29.92$\pm$0.77&15.01$\pm$0.58&18.11$\pm$0.53&13.41$\pm$0.68&12.48$\pm$0.6&45.47$\pm$0.72&16.42$\pm$0.57\\
090328713&90328B&--&--&2.91$\pm$0.34&0.26$\pm$0.19&0.48$\pm$0.25&1.89$\pm$0.39&0.29$\pm$0.19\\
090330279&090330&132.74$\pm$1.26&214.69$\pm$1.19&256.45$\pm$1.54&33.34$\pm$0.59&74.62$\pm$1.81&125.5$\pm$1.82&24.90$\pm$0.51\\
090331681&090331&179.36$\pm$1.27&138.34$\pm$1.16&207.42$\pm$1.47&214.46$\pm$1.52&--&179.81$\pm$1.7&0.38$\pm$0.37\\
090403314&090403&27.33$\pm$0.66&--&--&14.08$\pm$0.51&8.13$\pm$0.5&--&14.3$\pm$0.58\\
090405663&090405&--&2.08$\pm$0.3&0.67$\pm$0.3&--&--&--&0.64$\pm$0.33\\
090409288&090409&--&--&--&22.4$\pm$0.51&13.57$\pm$0.6&--&49.79$\pm$0.65\\
090411838&090411&18.18$\pm$0.56&15.42$\pm$0.57&15.65$\pm$0.54&18.85$\pm$0.74&16.29$\pm$0.6&--&15.81$\pm$0.47\\
090411991&090411B&36.54$\pm$0.89&16.99$\pm$0.56&15.3$\pm$0.54&14.02$\pm$0.69&14.94$\pm$0.54&--&17.34$\pm$0.43\\
090412061&090412&--&--&--&--&5.76$\pm$0.43&2.11$\pm$0.2&1.47$\pm$0.41\\
090418816&090418C&--&0.42$\pm$0.16&--&--&--&--&--\\
090422150&090422&--&--&13.34$\pm$0.54&1.38$\pm$0.52&17.73$\pm$0.52&--&--\\
090423330&090423&--&12.67$\pm$0.59&36.67$\pm$0.72&52.83$\pm$0.84&--&--&12.35$\pm$0.5\\
090424592&090424&57.38$\pm$0.8&46.21$\pm$0.64&34.46$\pm$0.8&4.83$\pm$0.23&24.7$\pm$0.96&3.94$\pm$0.52&16.52$\pm$0.32\\
090425377&090425&18.18$\pm$0.85&18.27$\pm$0.83&67.49$\pm$0.76&75.58$\pm$0.79&12.38$\pm$0.81&40.96$\pm$1.29&73.41$\pm$0.61\\
090426066&090426B&12.77$\pm$0.4&9.12$\pm$0.46&--&--&--&3.65$\pm$0.59&12.83$\pm$0.39\\
090426690&090426C&5.22$\pm$0.43&6.53$\pm$0.45&3.78$\pm$0.5&3.42$\pm$0.48&4.8$\pm$0.5&--&5.06$\pm$0.41\\
090427644&090427B&--&--&--&--&--&20.45$\pm$0.76&1.57$\pm$0.40\\
090427688&090427C&--&11.2$\pm$0.56&12.26$\pm$0.54&15.14$\pm$0.64&--&--&12.48$\pm$0.53\\
090428552&090428B &18.91$\pm$0.59&19.62$\pm$0.71&12.19$\pm$0.44&11.01$\pm$0.5&15.01$\pm$0.57&50.98$\pm$1.05&29.02$\pm$0.59\\
090429530&090429C&18.02$\pm$0.48&14.14$\pm$0.46&--&10.34$\pm$0.44&9.98$\pm$0.56&13.54$\pm$0.53&10.24$\pm$0.36\\
090429753&090429D&--&--&2.4$\pm$0.48&0.61$\pm$0.34&7.68$\pm$0.51&53.34$\pm$1.04&1.98$\pm$0.45\\
090502777&090502&56.1$\pm$0.85&22.21$\pm$0.83&31.49$\pm$0.92&54.66$\pm$0.66&--&32.77$\pm$1.1&59.78$\pm$0.73\\
090509215&090509&141.28$\pm$1.69&36.61$\pm$0.57&247.14$\pm$1.23&31.97$\pm$0.74&26.21$\pm$0.64&120.93$\pm$1.97&261.18$\pm$1.08\\
090510016&090510&0.48$\pm$0.25&0.38$\pm$0.14&0.64$\pm$0.19&0.42$\pm$0.12&0.35$\pm$0.07&0.38$\pm$0.14&0.38$\pm$0.05\\
090510325&090510B&--&29.12$\pm$0.63&28.29$\pm$0.61&18.53$\pm$0.36&7.2$\pm$0.39&--&10.59$\pm$0.46\\
090511684&090511&--&36.83$\pm$0.63&4.48$\pm$0.45&1.66$\pm$0.45&9.95$\pm$0.52&--&5.22$\pm$0.43\\
090513916&090513&28.93$\pm$0.79&--&--&24.16$\pm$0.68&18.66$\pm$0.68&--&18.46$\pm$0.52\\
090513941&090513B&--&--&46.11$\pm$0.72&17.41$\pm$0.59&13.95$\pm$0.5&44.42$\pm$0.75&12.74$\pm$0.46\\
090514006&090514&43.97$\pm$0.73&42.78$\pm$0.66&50.37$\pm$0.71&43.1$\pm$0.62&31.23$\pm$0.57&--&45.28$\pm$0.54\\
090516137&090516B&--&97.02$\pm$1.22&--&--&--&--&126.27$\pm$1.59\\
090516353&090516&95.62$\pm$1.02&93.5$\pm$0.81&89.34$\pm$0.66&84.58$\pm$0.86&102.69$\pm$1&--&93.38$\pm$0.49\\
090516853&090516C&16.13$\pm$0.63&17.95$\pm$0.68&13.28$\pm$0.44&11.17$\pm$0.59&10.78$\pm$0.6&24.67$\pm$0.79&13.98$\pm$0.39\\
090518080&090518&50.72$\pm$2.29&75.42$\pm$0.99&20.13$\pm$0.37&38.66$\pm$0.73&8.32$\pm$0.5&55.58$\pm$0.84&7.97$\pm$0.44\\
090518244&090518B&10.88$\pm$0.73&9.15$\pm$0.6&8.03$\pm$0.57&6.75$\pm$0.52&5.98$\pm$0.49&--&6.62$\pm$0.4\\
090519462&090519B&139.52$\pm$1.3&--&--&--&12.1$\pm$0.46&--&37.79$\pm$0.50\\
090519881&090519&45.86$\pm$1.56&87.97$\pm$1.53&--&14.98$\pm$0.56&252.58$\pm$1.53&--&42.88$\pm$0.57\\
090520832&090520B&20.61$\pm$0.68&--&--&--&--&54.08$\pm$1.25&--\\
090520850&090520C&8.48$\pm$0.43&7.01$\pm$0.4&4.22$\pm$0.33&3.81$\pm$0.25&4.1$\pm$0.37&4.16$\pm$0.64&4.03$\pm$0.23\\
090520876&090520D&17.15$\pm$0.72&14.62$\pm$0.64&15.58$\pm$0.56&11.65$\pm$0.43&18.98$\pm$0.72&43.65$\pm$0.73&14.37$\pm$0.5\\
090522344&090522&--&28.06$\pm$0.62&11.97$\pm$0.54&8.16$\pm$0.53&--&44.99$\pm$1.12&19.26$\pm$0.57\\
090524346&090524&46.88$\pm$0.7&52.29$\pm$0.72&50.21$\pm$0.66&48.42$\pm$0.68&46.46$\pm$0.57&67.01$\pm$1.37&50.46$\pm$0.58\\
090528173&090528&33.95$\pm$0.8&32.13$\pm$0.66&33.18$\pm$0.62&27.46$\pm$0.63&25.82$\pm$0.69&--&29.06$\pm$0.55\\
090528516&090528B&106.21$\pm$0.92&100.54$\pm$0.89&91.52$\pm$0.87&76.48$\pm$0.93&44.38$\pm$0.69&--&92.1$\pm$0.73\\
090529310&090529B&25.09$\pm$0.57&4.8$\pm$0.54&21.79$\pm$0.59&4.83$\pm$0.48&45.31$\pm$0.89&5.92$\pm$0.84&4.29$\pm$0.45\\
090529564&090529C&12.9$\pm$0.69&10.21$\pm$0.39&10.4$\pm$0.34&9.86$\pm$0.23&9.63$\pm$0.25&38.98$\pm$0.82&10.14$\pm$0.25\\
090530760&090530B&147.42$\pm$1.33&143.97$\pm$1.09&131.36$\pm$1.27&114.34$\pm$1.14&89.76$\pm$1.07&175.68$\pm$1.72&135.68$\pm$0.91\\
090531775&90531B&--&--&23.84$\pm$0.56&2.5$\pm$0.54&2.02$\pm$0.48&14.94$\pm$0.58&2.02$\pm$0.39\\
090602564&090602&47.36$\pm$0.75&--&--&10.53$\pm$0.52&13.79$\pm$0.61&44.22$\pm$0.61&13.66$\pm$0.49\\
090606471&090606&--&96.64$\pm$1.03&--&37.57$\pm$0.98&41.44$\pm$1.39&64.26$\pm$1.25&--\\
090608052&090608&24.32$\pm$1&9.82$\pm$0.58&10.14$\pm$0.48&30.37$\pm$0.6&22.11$\pm$0.46&--&8.22$\pm$0.35\\
090610723&090610B&179.74$\pm$1.19&196.93$\pm$1.54&100.61$\pm$1.39&142.43$\pm$1.23&122.24$\pm$1.59&46.05$\pm$1.61&140.74$\pm$0.83\\
090610883&090610C&13.15$\pm$0.42&13.18$\pm$0.39&13.15$\pm$0.42&13.15$\pm$0.42&13.15$\pm$0.42&13.18$\pm$0.39&13.15$\pm$0.42\\
090612619&090612&34.3$\pm$0.59&4.19$\pm$0.57&5.18$\pm$0.5&4.99$\pm$0.42&5.79$\pm$0.58&30.88$\pm$0.9&7.23$\pm$0.55\\
090616157&090616&--&--&26.43$\pm$0.76&5.76$\pm$0.73&--&--&2.08$\pm$0.61\\
090617208&090617&2.14$\pm$0.34&--&--&0.99$\pm$0.25&0.22$\pm$0.07&1.47$\pm$0.29&0.26$\pm$0.14\\
090618353&090618&169.25$\pm$1.43&118.88$\pm$1.15&130.53$\pm$1.3&107.58$\pm$1.06&104.83$\pm$0.93&25.47$\pm$0.99&130.24$\pm$1.05\\
090620400&090620&18.98$\pm$0.79&10.75$\pm$0.6&13.95$\pm$0.55&10.91$\pm$0.52&7.3$\pm$0.37&41.98$\pm$0.96&14.34$\pm$0.47\\
090621185&090621&50.62$\pm$0.87&61.86$\pm$1.02&67.33$\pm$1.03&44.16$\pm$0.89&51.87$\pm$0.92&85.47$\pm$1.13&63.71$\pm$0.9\\
090621417&090621C&47.04$\pm$0.69&41.66$\pm$0.73&35.17$\pm$0.69&30.75$\pm$0.64&23.23$\pm$0.63&--&36.42$\pm$0.57\\
090621447&090621D&38.11$\pm$0.58&23.23$\pm$0.72&28.99$\pm$0.6&20.1$\pm$0.56&21.06$\pm$0.6&181.44$\pm$1.68&21.63$\pm$0.53\\
090621922&090621B &--&3.33$\pm$0.28&--&0.29$\pm$0.16&0.19$\pm$0.1&--&0.29$\pm$0.16\\
090623107&090623&79.9$\pm$1&41.44$\pm$0.73&44.29$\pm$0.76&52.1$\pm$0.73&87.46$\pm$1.01&129.06$\pm$1.15&62.59$\pm$0.75\\
090625234&090625&--&--&23.2$\pm$0.49&13.86$\pm$0.48&10.08$\pm$0.48&187.26$\pm$1.68&16.26$\pm$0.46\\
090625560&090625B&40.26$\pm$0.96&88.99$\pm$0.92&23.36$\pm$0.92&18.43$\pm$0.6&19.94$\pm$0.59&--&6.85$\pm$0.45\\
090626189&090626&53.5$\pm$0.72&48$\pm$0.59&45.31$\pm$0.33&45.18$\pm$0.33&44.54$\pm$0.45&2.66$\pm$0.39&47.07$\pm$0.55\\
090630311&090630&22.98$\pm$0.37&11.55$\pm$0.67&3.46$\pm$0.45&2.94$\pm$0.55&23.52$\pm$0.5&31.52$\pm$0.85&10.08$\pm$0.52\\
090701225&090701&--&12.48$\pm$0.73&--&19.33$\pm$0.64&--&37.41$\pm$0.66&--\\
090703329&090703&28.8$\pm$0.69&15.68$\pm$0.6&22.72$\pm$0.49&6.56$\pm$0.48&40.99$\pm$0.7&2.08$\pm$0.84&9.15$\pm$0.55\\
090704242&090704&54.43$\pm$1.05&60.99$\pm$0.8&58.82$\pm$0.85&59.71$\pm$0.67&48.54$\pm$0.72&--&58.69$\pm$0.62\\
090706283&090706&--&--&--&18.69$\pm$0.37&--&--&--\\
090708152&090708&15.52$\pm$0.49&42.82$\pm$0.85&7.84$\pm$0.48&34.69$\pm$0.61&31.84$\pm$0.69&--&12.48$\pm$0.39\\
090709630&090709B&79.36$\pm$0.95&4.06$\pm$0.39&11.97$\pm$0.47&9.7$\pm$0.39&12.19$\pm$0.49&96.8$\pm$1.11&11.14$\pm$0.32\\
090712160&090712&--&--&53.47$\pm$0.8&51.78$\pm$0.79&31.94$\pm$0.55&55.36$\pm$0.87&31.68$\pm$0.62\\
090713020&090713&--&50.02$\pm$0.8&47.87$\pm$0.8&43.2$\pm$0.79&153.63$\pm$1.35&131.9$\pm$1.35&49.86$\pm$0.73\\
090717034&090717&--&--&--&10.46$\pm$0.53&55.65$\pm$0.94&76.58$\pm$1.25&65.63$\pm$0.93\\
090717111&090717B&--&--&--&0.67$\pm$0.25&0.58$\pm$0.23&1.44$\pm$0.39&--\\
090718720&090718&90.21$\pm$1.22&15.1$\pm$1.04&--&--&--&--&--\\
090718762&090718B&22.02$\pm$0.73&24.74$\pm$0.86&20.16$\pm$0.68&21.47$\pm$0.67&17.92$\pm$0.55&--&21.41$\pm$0.48\\
090719063&090719&33.28$\pm$0.57&12.06$\pm$0.76&12.74$\pm$0.6&10.85$\pm$0.48&9.02$\pm$0.47&11.71$\pm$0.75&11.49$\pm$0.4\\
090720276&090720&5.34$\pm$0.52&5.5$\pm$0.51&5.89$\pm$0.51&3.23$\pm$0.39&5.09$\pm$0.57&--&5.6$\pm$0.4\\
090720710&090720B&--&--&6.46$\pm$0.54&8.06$\pm$0.49&7.33$\pm$0.39&9.25$\pm$0.54&8.64$\pm$0.51\\
090802235&090802&--&--&--&--&0.19$\pm$0.5&2.4$\pm$0.58&--\\
090802666&090802B&--&19.78$\pm$0.7&19.46$\pm$0.51&13.89$\pm$0.57&55.68$\pm$0.89&--&18.21$\pm$0.53\\
090807832&090807B&5.63$\pm$0.59&6.34$\pm$0.5&1.73$\pm$0.42&1.25$\pm$0.57&--&--&45.79$\pm$1.34\\
090809978&090809B&13.34$\pm$0.7&12.7$\pm$0.52&10.08$\pm$0.49&9.5$\pm$0.52&12.9$\pm$0.66&2.85$\pm$0.58&13.22$\pm$0.52\\
090813174&090813&10.27$\pm$0.66&9.12$\pm$0.61&13.66$\pm$0.61&2.5$\pm$0.51&10.82$\pm$0.7&47.42$\pm$0.93&8.96$\pm$0.55\\
090814368&090814C&--&--&--&1.31$\pm$0.25&0.26$\pm$0.14&--&0.51$\pm$0.18\\
090815438&090815B&23.04$\pm$0.65&26.5$\pm$0.6&24.26$\pm$0.65&23.65$\pm$0.66&12.16$\pm$0.72&82.91$\pm$1.41&25.28$\pm$0.52\\
090817036&090817&75.9$\pm$0.88&89.44$\pm$1.92&33.25$\pm$0.73&10.37$\pm$0.52&12.99$\pm$0.53&259.87$\pm$1.65&33.47$\pm$0.59\\
090820027&090820&6.18$\pm$0.3&5.76$\pm$0.18&4.86$\pm$0.09&6.18$\pm$0.12&6.08$\pm$0.09&6.62$\pm$0.66&6.24$\pm$0.07\\
090820509&090820B&20.38$\pm$0.58&11.23$\pm$0.59&12.06$\pm$0.61&1.86$\pm$0.59&11.97$\pm$0.82&40.9$\pm$0.89&11.87$\pm$0.49\\
090826068&090826&28.8$\pm$0.66&25.44$\pm$0.69&--&16.83$\pm$0.72&20.13$\pm$0.56&--&10.02$\pm$0.43\\
090828099&090828&23.42$\pm$1.09&23.78$\pm$0.84&19.23$\pm$0.76&19.58$\pm$0.87&16.48$\pm$0.77&120.32$\pm$1.32&88.96$\pm$0.97\\
090829672&090829&73.22$\pm$1.15&70.85$\pm$0.93&74.72$\pm$0.78&62.66$\pm$1.03&49.02$\pm$1&145.79$\pm$1.22&77.79$\pm$0.78\\
090829702&090829B&55.36$\pm$0.81&26.69$\pm$0.57&18.21$\pm$0.49&18.37$\pm$0.47&92.8$\pm$0.92&--&129.02$\pm$0.92\\
090831317&090831&59.84$\pm$0.96&54.14$\pm$0.94&40.86$\pm$0.7&157.18$\pm$1.23&41.18$\pm$0.62&152.83$\pm$1.68&58.34$\pm$0.68\\
090902401&090902&11.3$\pm$0.63&--&11.01$\pm$0.51&1.15$\pm$0.27&2.62$\pm$0.34&10.75$\pm$0.55&1.28$\pm$0.28\\
090904058&090904B&50.53$\pm$0.89&51.87$\pm$0.79&53.15$\pm$0.72&52.16$\pm$0.75&53.34$\pm$0.77&82.21$\pm$1.2&52.35$\pm$0.59\\
090910812&090910&38.53$\pm$0.69&40.38$\pm$0.52&40.1$\pm$0.44&121.31$\pm$1.02&38.91$\pm$0.73&--&40.19$\pm$0.23\\
090912660&090912&262.5$\pm$1.53&55.94$\pm$1.06&169.15$\pm$0.87&120.03$\pm$0.9&9.06$\pm$0.39&--&126.37$\pm$0.54\\
090922539&090922A&96.58$\pm$1.09&88.48$\pm$0.9&87.17$\pm$0.83&88.03$\pm$0.95&7.17$\pm$0.46&100.7$\pm$1.02&88.42$\pm$0.68\\
090926181&090926&13.92$\pm$0.42&19.84$\pm$0.74&13.5$\pm$0.2&12.29$\pm$0.45&11.39$\pm$0.2&10.34$\pm$0.62&17.63$\pm$0.55\\
090926914&090926B&31.1$\pm$0.74&32.13$\pm$0.79&35.26$\pm$0.75&41.82$\pm$0.72&35.17$\pm$0.7&29.89$\pm$1.27&41.34$\pm$0.62\\
090927422&090927&2.27$\pm$0.48&--&--&1.25$\pm$0.44&--&--&3.2$\pm$0.32\\
090929190&090929A&53.38$\pm$1.08&8.74$\pm$0.52&4.8$\pm$0.36&5.89$\pm$0.4&4.29$\pm$0.29&2.21$\pm$0.48&7.81$\pm$0.45\\
091003191&091003&21.22$\pm$0.58&20.03$\pm$0.51&21.34$\pm$0.52&20.58$\pm$0.39&20.67$\pm$0.39&18.78$\pm$0.66&23.17$\pm$0.57\\
091010113&091010&8.64$\pm$0.49&7.65$\pm$0.32&6.94$\pm$0.27&6.18$\pm$0.27&5.89$\pm$0.33&--&7.58$\pm$0.3\\
091020900&091020&11.84$\pm$0.54&47.42$\pm$0.87&12.67$\pm$0.52&16.83$\pm$0.7&30.5$\pm$0.78&--&44.67$\pm$0.65\\
091024372&091024&70.53$\pm$0.87&22.53$\pm$0.57&66.02$\pm$0.82&49.63$\pm$0.84&44.1$\pm$0.59&--&48.26$\pm$0.62\\
091026550&091026&14.14$\pm$0.55&--&--&24.99$\pm$0.74&6.88$\pm$0.43&28.77$\pm$0.91&15.74$\pm$0.46\\
091030828&091030&26.08$\pm$0.84&143.17$\pm$1.32&36.54$\pm$0.59&99.55$\pm$0.91&35.65$\pm$0.57&19.52$\pm$0.83&97.63$\pm$0.81\\
091031500&091031&11.49$\pm$0.57&37.41$\pm$0.59&16.35$\pm$0.64&16.13$\pm$0.51&15.97$\pm$0.43&20.83$\pm$0.73&37.86$\pm$0.66\\
091102607&091102&--&7.2$\pm$0.57&6.78$\pm$0.5&7.49$\pm$0.56&6.53$\pm$0.5&--&7.58$\pm$0.46\\
091112737&091112&68.03$\pm$0.95&36.74$\pm$0.72&19.49$\pm$0.68&15.39$\pm$0.63&15.49$\pm$0.51&58.5$\pm$0.97&50.85$\pm$0.82\\
091120191&091120&54.46$\pm$0.82&50.91$\pm$0.64&50.72$\pm$0.58&51.39$\pm$0.53&51.62$\pm$0.66&--&50.94$\pm$0.43\\
091123298&091123&55.71$\pm$1.75&203.04$\pm$1.41&184.13$\pm$1.09&174.98$\pm$1.09&177.95$\pm$1.13&86.05$\pm$2.07&200.61$\pm$1.14\\
091126333&091126&--&--&1.06$\pm$0.34&0.29$\pm$0.16&0.29$\pm$0.17&1.6$\pm$0.42&0.32$\pm$0.18\\
091126389&091126B&--&--&--&--&--&--&0.86$\pm$0.3\\
091127976&091127&10.24$\pm$0.43&9.63$\pm$0.42&8.45$\pm$0.33&7.71$\pm$0.3&5.92$\pm$0.5&20.99$\pm$0.74&9.15$\pm$0.26\\
091128285&091128&72.67$\pm$1.02&51.9$\pm$0.73&38.18$\pm$0.62&39.07$\pm$0.66&30.05$\pm$0.7&--&37.82$\pm$0.53\\
091208410&091208B&10.27$\pm$0.32&11.81$\pm$0.26&11.9$\pm$0.24&12.7$\pm$0.36&2.62$\pm$0.36&--&11.39$\pm$0.14\\
091221870&091221&68.93$\pm$0.9&24.32$\pm$0.77&80.58$\pm$1.02&30.78$\pm$0.66&16.03$\pm$0.52&--&30.43$\pm$0.62\\
100111176&100111A&--&46.66$\pm$0.81&41.63$\pm$0.76&12.54$\pm$0.65&--&47.49$\pm$0.79&11.01$\pm$0.46\\
100116897&100116A&19.17$\pm$1.15&15.97$\pm$0.97&100.06$\pm$0.9&16.83$\pm$0.95&102.21$\pm$0.87&7.78$\pm$1.06&100.93$\pm$0.71\\
100117879&100117A&--&--&--&0.48$\pm$0.21&0.51$\pm$0.32&2.34$\pm$0.26&0.51$\pm$0.19\\
100122616&100122A&8.13$\pm$0.43&9.15$\pm$0.45&7.65$\pm$0.36&6.3$\pm$0.37&5.34$\pm$0.4&--&7.94$\pm$0.33\\
100130729&100130A&36.26$\pm$1.07&106.4$\pm$0.98&96.67$\pm$0.82&93.5$\pm$0.82&23.55$\pm$1.06&--&93.92$\pm$0.68\\
100130777&100130B&58.66$\pm$0.9&41.47$\pm$0.74&69.41$\pm$0.95&44.32$\pm$0.63&42.62$\pm$0.82&137.18$\pm$1.16&71.07$\pm$0.76\\
100131730&100131A&3.62$\pm$0.39&3.68$\pm$0.3&3.49$\pm$0.2&3.26$\pm$0.19&3.14$\pm$0.28&2.4$\pm$0.48&3.14$\pm$0.14\\
100205490&100205B&31.68$\pm$0.61&21.18$\pm$0.68&36.9$\pm$0.72&8.67$\pm$0.53&15.52$\pm$0.73&--&14.3$\pm$0.44\\
100206563&100206A&--&1.76$\pm$0.39&0.48$\pm$0.26&1.18$\pm$0.12&0.22$\pm$0.16&--&0.19$\pm$0.13\\
100212588&100212A&5.82$\pm$0.41&11.33$\pm$0.46&17.44$\pm$0.48&--&--&9.18$\pm$0.49&8.1$\pm$0.46\\
100218194&100218A&14.53$\pm$1&--&57.28$\pm$0.83&30.27$\pm$0.72&36.1$\pm$0.63&--&31.33$\pm$0.7\\
100223110&100223A&--&--&--&0.51$\pm$0.19&0.26$\pm$0.09&0.22$\pm$0.16&0.26$\pm$0.1\\
100224112&100224B&33.34$\pm$0.61&87.3$\pm$0.93&77.98$\pm$0.82&74.18$\pm$0.91&113.38$\pm$1.13&27.01$\pm$1.16&76.42$\pm$0.81\\
100225115&100225A&68.77$\pm$0.99&34.91$\pm$0.73&11.26$\pm$0.45&19.74$\pm$0.57&13.34$\pm$0.58&20.48$\pm$0.56&13.02$\pm$0.44\\
100322045&100322A&38.21$\pm$0.79&36.29$\pm$0.45&36.64$\pm$0.36&36.48$\pm$0.33&36.93$\pm$0.28&10.43$\pm$0.77&36.83$\pm$0.21\\
100325275&100325A&9.06$\pm$0.39&7.81$\pm$0.41&15.62$\pm$0.57&6.98$\pm$0.32&6.78$\pm$0.32&2.78$\pm$0.52&7.52$\pm$0.25\\
100401297&100401A&181.73$\pm$2.18&127.68$\pm$1.17&--&91.87$\pm$1.5&--&--&89.06$\pm$1.32\\
100413732&100413A&204.99$\pm$1.57&173.95$\pm$1.46&213.57$\pm$1.17&170.46$\pm$1.23&62.3$\pm$1.25&192.74$\pm$2.09&120.06$\pm$1.00\\
100414097&100414A&23.52$\pm$0.66&24.45$\pm$0.6&47.3$\pm$0.79&44.03$\pm$1.09&22.05$\pm$0.3&21.31$\pm$0.54&22.53$\pm$0.32\\
100423244&100423B&47.9$\pm$0.64&63.23$\pm$0.79&21.98$\pm$0.66&21.09$\pm$0.67&11.97$\pm$0.55&23.46$\pm$0.84&19.84$\pm$0.55\\
100427356&100427A&26.08$\pm$0.68&7.14$\pm$0.52&9.38$\pm$0.58&9.79$\pm$0.56&10.56$\pm$0.55&--&10.43$\pm$0.41\\
100503554&100503A&102.85$\pm$0.96&20.45$\pm$0.89&149.47$\pm$1.31&135.97$\pm$1.2&24.64$\pm$0.8&--&147.58$\pm$1.09\\
100504806&100504A&52.67$\pm$0.92&22.75$\pm$0.59&19.55$\pm$0.73&22.82$\pm$0.68&12.93$\pm$0.6&--&24.51$\pm$0.73\\
100510810&100510A&45.44$\pm$0.83&18.14$\pm$0.58&21.06$\pm$0.56&16.74$\pm$0.54&18.78$\pm$0.68&65.86$\pm$0.83&19.81$\pm$0.54\\
100511035&100511A&31.94$\pm$0.7&33.34$\pm$0.61&44.7$\pm$0.78&43.04$\pm$0.66&29.63$\pm$0.78&15.3$\pm$0.69&41.41$\pm$0.67\\
100522157&100522A&36.42$\pm$0.67&36.58$\pm$0.62&37.28$\pm$0.56&3.26$\pm$0.32&3.1$\pm$0.34&--&37.38$\pm$0.59\\
100528075&100528A&16.99$\pm$0.53&24.93$\pm$0.62&19.01$\pm$0.66&15.58$\pm$0.6&14.72$\pm$0.59&18.59$\pm$0.95&22.98$\pm$0.67\\
100615083&100615A&34.18$\pm$0.75&36$\pm$0.72&39.3$\pm$0.99&37.47$\pm$0.84&34.05$\pm$0.75&65.31$\pm$1.13&36.42$\pm$0.59\\
100619015&100619A&96.22$\pm$0.92&12.83$\pm$1.02&95.04$\pm$0.73&92.61$\pm$0.71&92.22$\pm$0.85&--&92.61$\pm$0.34\\
100625773&100625A&--&0.64$\pm$0.3&--&1.63$\pm$0.36&0.26$\pm$0.05&--&0.38$\pm$0.14\\
100701490&100701B&52.22$\pm$0.69&11.36$\pm$0.57&7.58$\pm$0.52&15.33$\pm$0.64&22.59$\pm$0.81&8.1$\pm$0.49&25.02$\pm$0.63\\
100704149&100704A&141.06$\pm$1.82&262.11$\pm$1.55&14.56$\pm$0.59&12.19$\pm$0.6&11.14$\pm$0.56&152.13$\pm$1.64&183.62$\pm$1.12\\
100707032&100707A&70.75$\pm$0.81&--&17.98$\pm$0.43&14.62$\pm$0.4&11.9$\pm$0.33&6.75$\pm$0.56&15.55$\pm$0.43\\
100722096&100722A&7.74$\pm$0.46&8.16$\pm$0.46&7.3$\pm$0.34&9.22$\pm$0.73&3.81$\pm$0.46&--&7.36$\pm$0.28\\
100724029&100724B&118.14$\pm$1.03&119.04$\pm$0.69&118.53$\pm$0.75&116$\pm$0.95&104.54$\pm$1.14&68.8$\pm$0.69&116.51$\pm$0.56\\
100727238&100727A&23.17$\pm$0.53&26.91$\pm$0.54&22.18$\pm$0.48&42.46$\pm$0.76&--&--&25.92$\pm$0.47\\
100728095&100728A&118.46$\pm$1.06&147.1$\pm$1.06&157.95$\pm$1&157.82$\pm$1.08&158.34$\pm$1.52&45.09$\pm$1.04&159.97$\pm$0.76\\
100728439&100728B&5.28$\pm$0.52&12.54$\pm$0.55&9.66$\pm$0.51&8.67$\pm$0.52&9.15$\pm$0.55&--&9.34$\pm$0.46\\
100802240&100802A&134.3$\pm$0.94&273.12$\pm$1.59&182.4$\pm$1.24&131.71$\pm$1.12&10.14$\pm$0.43&180.48$\pm$1.2&132.26$\pm$0.84\\
100814160&100814A&154.21$\pm$0.92&76.83$\pm$0.7&138.21$\pm$0.92&133.41$\pm$0.92&16.32$\pm$0.52&173.98$\pm$1.87&25.41$\pm$0.29\\
100814351&100814B&8.74$\pm$0.44&5.06$\pm$0.27&4.19$\pm$0.34&6.34$\pm$0.37&5.98$\pm$0.34&--&5.18$\pm$0.32\\
100816026&100816A&2.53$\pm$0.39&2.14$\pm$0.3&2.5$\pm$0.34&2.3$\pm$0.28&2.18$\pm$0.36&--&2.24$\pm$0.23\\
100906576&100906A&118.11$\pm$0.9&119.33$\pm$1&109.92$\pm$0.87&16.1$\pm$0.56&11.84$\pm$0.29&100.45$\pm$1.3&115.55$\pm$0.68\\
100910818&100910A&13.44$\pm$0.47&13.28$\pm$0.32&13.06$\pm$0.23&12.1$\pm$0.28&8.64$\pm$0.24&--&13.38$\pm$0.29\\
100915243&100915B&--&58.85$\pm$0.72&51.68$\pm$0.86&12.42$\pm$0.45&8.8$\pm$0.39&--&5.76$\pm$0.45\\
100924165&100924A&11.17$\pm$0.44&13.02$\pm$0.49&15.78$\pm$0.66&7.04$\pm$0.32&2.91$\pm$0.39&--&11.87$\pm$0.44\\
101008697&101008A&--&18.53$\pm$0.53&22.66$\pm$0.71&14.37$\pm$0.54&14.02$\pm$0.52&--&7.2$\pm$0.3\\
101011707&101011A&--&109.18$\pm$1.05&--&102.37$\pm$1.1&34.62$\pm$0.62&--&38.5$\pm$0.62\\
101014175&101014A&207.81$\pm$0.47&208.54$\pm$0.37&210.75$\pm$0.39&214.14$\pm$0.58&219.23$\pm$1.02&249.63$\pm$1.77&211.17$\pm$0.16\\
101023951&101023A&24.42$\pm$0.84&72.8$\pm$0.83&75.78$\pm$0.81&75.42$\pm$0.88&71.1$\pm$0.94&10.69$\pm$1.18&66.94$\pm$0.55\\
101024486&101024A&--&1.06$\pm$0.57&--&1.47$\pm$0.46&11.49$\pm$0.57&--&20.51$\pm$0.54\\
101112924&101112A&3.84$\pm$0.45&6.88$\pm$0.43&15.3$\pm$0.4&3.94$\pm$0.49&6.18$\pm$0.48&29.89$\pm$0.8&3.97$\pm$0.32\\
101123952&101123A&102.56$\pm$0.8&103.07$\pm$0.7&103.94$\pm$0.67&102.37$\pm$0.56&101.41$\pm$0.6&17.25$\pm$0.92&103.01$\pm$0.44\\
101129652&101129A&--&--&--&0.67$\pm$0.25&0.54$\pm$0.23&--&0.54$\pm$0.17\\
101201418&101201A&68$\pm$0.87&63.87$\pm$0.76&64.64$\pm$0.77&56.96$\pm$0.79&62.05$\pm$1&33.73$\pm$1.23&83.62$\pm$0.68\\
101213451&101213A&51.71$\pm$0.76&50.56$\pm$0.77&28.29$\pm$0.6&23.84$\pm$0.48&24.7$\pm$0.43&81.47$\pm$0.96&38.24$\pm$0.56\\
101219686&101219B&--&--&32.1$\pm$0.54&58.05$\pm$0.83&20.64$\pm$0.49&--&54.21$\pm$0.75\\
101224227&101224A&--&--&0.61$\pm$0.3&0.45$\pm$0.19&--&--&0.58$\pm$0.32\\
110102788&110102A&128.99$\pm$0.74&137.5$\pm$0.68&136.77$\pm$0.75&132.93$\pm$0.81&132.45$\pm$0.74&--&133.76$\pm$0.34\\
110106893&110106B&24.54$\pm$0.76&29.79$\pm$0.63&27.36$\pm$0.63&24.64$\pm$0.7&34.62$\pm$0.67&--&22.78$\pm$0.62\\
110112934&110112B&--&1.66$\pm$0.3&--&--&0.35$\pm$0.2&--&0.26$\pm$0.09\\
110119931&110119A&258.37$\pm$1.47&28.16$\pm$0.91&67.33$\pm$0.63&58.82$\pm$0.76&60.93$\pm$0.81&280.74$\pm$1.88&59.87$\pm$0.59\\
110120666&110120A&--&3.84$\pm$0.54&9.82$\pm$0.52&22.4$\pm$0.59&13.06$\pm$0.43&22.53$\pm$0.67&16.06$\pm$0.39\\
110123804&110123A&16.45$\pm$0.55&20.22$\pm$0.63&19.1$\pm$0.5&17.15$\pm$0.52&15.94$\pm$0.46&46.94$\pm$0.63&17.18$\pm$0.42\\
110125894&110125A&4.77$\pm$0.39&3.81$\pm$0.39&2.5$\pm$0.45&5.02$\pm$0.39&6.5$\pm$0.44&--&4.1$\pm$0.32\\
110128073&110128A&69.5$\pm$1.12&--&12.77$\pm$0.79&--&70.5$\pm$0.74&67.78$\pm$0.85&--\\
110201399&110201A&--&--&42.78$\pm$0.6&13.41$\pm$0.53&4.29$\pm$0.36&--&11.71$\pm$0.51\\
110207470&110207A&12.06$\pm$0.84&18.4$\pm$0.66&47.26$\pm$0.8&38.08$\pm$0.73&0.51$\pm$0.29&43.71$\pm$0.87&39.01$\pm$0.52\\
110213220&110213A&35.81$\pm$0.63&33.06$\pm$0.64&30.98$\pm$0.57&32.58$\pm$0.67&14.56$\pm$0.67&29.54$\pm$1.11&33.12$\pm$0.46\\
110301214&110301A&7.55$\pm$0.33&7.14$\pm$0.3&5.95$\pm$0.2&5.38$\pm$0.23&5.06$\pm$0.16&9.76$\pm$0.6&6.14$\pm$0.13\\
110318552&110318A&13.79$\pm$0.76&15.68$\pm$0.5&9.95$\pm$0.54&11.62$\pm$0.75&5.22$\pm$0.48&46.4$\pm$0.82&14.66$\pm$0.41\\
110319815&110319B&--&--&49.12$\pm$0.8&--&53.5$\pm$0.69&--&13.95$\pm$0.52\\
110328520&110328B&49.7$\pm$0.79&82.75$\pm$0.87&44.03$\pm$0.77&77.6$\pm$0.9&93.89$\pm$1.13&29.89$\pm$0.72&71.55$\pm$0.82\\
110401920&110401A&4.51$\pm$0.53&20.83$\pm$0.5&3.39$\pm$0.32&1.06$\pm$0.25&0.96$\pm$0.23&--&2.82$\pm$0.28\\
110402009&110402A&--&68.19$\pm$0.66&72.38$\pm$0.76&33.18$\pm$0.68&33.79$\pm$0.75&--&35.17$\pm$0.53\\
110412315&110412A&21.31$\pm$0.7&25.73$\pm$0.65&18.11$\pm$0.55&15.74$\pm$0.6&10.53$\pm$0.52&--&17.38$\pm$0.53\\
110420946&110420B&--&--&0.7$\pm$0.28&0.35$\pm$0.17&--&--&0.74$\pm$0.25\\
110529034&110529A&0.93$\pm$0.3&0.83$\pm$0.33&8.19$\pm$0.3&1.12$\pm$0.25&0.58$\pm$0.24&0.19$\pm$0.2&0.64$\pm$0.19\\
110610640&110610A&48.35$\pm$0.92&45.02$\pm$0.63&49.25$\pm$0.58&42.34$\pm$0.68&40.35$\pm$0.7&--&39.94$\pm$0.55\\
110625881&110625A&37.98$\pm$0.8&49.31$\pm$0.8&36.13$\pm$0.92&27.33$\pm$0.63&25.18$\pm$0.26&42.21$\pm$0.87&30.75$\pm$0.71\\
110705151&110705A&--&0.7$\pm$0.23&0.35$\pm$0.17&0.29$\pm$0.13&0.35$\pm$0.07&0.19$\pm$0.1&0.35$\pm$0.07\\
110709642&110709A&44.74$\pm$0.7&43.26$\pm$0.76&42.66$\pm$0.64&39.58$\pm$0.7&36.06$\pm$0.52&54.78$\pm$0.89&42.11$\pm$0.67\\
110721200&110721A&10.88$\pm$0.47&12.64$\pm$0.56&57.82$\pm$0.81&16.61$\pm$0.62&15.62$\pm$0.64&11.23$\pm$0.74&17.73$\pm$0.45\\
110731465&110731A&8.9$\pm$0.47&8.03$\pm$0.46&8.26$\pm$0.4&6.94$\pm$0.2&6.56$\pm$0.16&8.77$\pm$0.59&6.98$\pm$0.14\\
110818860&110818A&73.6$\pm$1.07&73.44$\pm$0.96&222.88$\pm$1.48&53.95$\pm$0.83&55.84$\pm$0.94&89.12$\pm$1.34&62.88$\pm$0.82\\
110825102&110825A&69.86$\pm$0.82&68.61$\pm$0.75&64.42$\pm$0.54&7.26$\pm$0.16&6.56$\pm$0.2&5.66$\pm$0.57&67.2$\pm$0.61\\
110903111&110903A&233.34$\pm$1.75&239.52$\pm$1.35&228.19$\pm$1.19&215.46$\pm$1.08&220.1$\pm$1.16&249.63$\pm$1.59&227.81$\pm$1.02\\
110921577&110921A&107.14$\pm$0.92&23.2$\pm$0.52&23.97$\pm$0.54&15.55$\pm$0.62&17.76$\pm$0.52&84.61$\pm$0.93&15.01$\pm$0.44\\
\enddata
$^{*}$ Available in the electronic version only.
\end{deluxetable}

\newpage
\begin{table}[tbp]
\centering
\caption{The ratio of short to long GRB numbers with a division of $T_{90}=2$ seconds for samples observed with different GRB missions and for {\em Fermi}/GBM observations in some bands. The results of bimodal distribution test with the KMM algorithm are also reported.}
\begin{tabular}{lccc}
\hline
Instrument &Band (keV) & SGRB:LGRB &$p_{\rm KMM}$\\ \hline
{\em HETE-2}/FREGATE &6-80 &0:82 &0.32\\
{\em SWIFT}/BAT &15-150 &51:557 &$7.5\times 10^{-22}$\\
{\em BeppoSAX}/GRBM &40-700  &111:892 &$1.8\times 10^{-18}$\\
{\em Fermi}/GBM &8-1000  &39:253 &$1.0\times 10^{-11}$\\
{\em CGRO}/BATSE &50-300  &500:1541 &0\\
{\em INTEGRAL}/SPI-ACS &20-8000 &196:724 &$3.0\times 10^{-30}$\\ \hline
GBM-1&8-15&5:236&$2.25\times 10^{-2}$\\
GBM-2&15-25&13:237&$5.9\times 10^{-4}$\\
GBM-3&25-50&12:246&$1.1\times 10^{-5}$\\
GBM-4&50-100&35:248&$6.6\times 10^{-8}$\\
GBM-5&100-350&32:240&$3.4\times 10^{-11}$\\
GBM-6&350-1000&13:177&$1.2\times 10^{-5}$\\
\hline
\end{tabular}
\end{table}
\newpage
\clearpage \thispagestyle{empty} \setlength{\voffset}{-18mm}

\newpage
\clearpage \thispagestyle{empty} \setlength{\voffset}{-18mm}

\newpage
\clearpage \thispagestyle{empty} \setlength{\voffset}{-18mm}

\newpage
\clearpage \thispagestyle{empty} \setlength{\voffset}{-18mm}

\newpage
\clearpage \thispagestyle{empty} \setlength{\voffset}{-18mm}

\newpage

\clearpage \thispagestyle{empty} \setlength{\voffset}{-18mm}

\newpage
\clearpage \thispagestyle{empty} \setlength{\voffset}{-18mm}
\newpage

\newpage
\end{document}